\documentclass[12pt,letterpaper]{article}
\pdfoutput=1

\usepackage{amsmath,amssymb,calc, amsthm,bbm, epsfig,psfrag, mathtools,comment}

\usepackage{dsfont}
\usepackage{mathrsfs}
\usepackage{esvect}
\usepackage{graphicx, enumerate, enumitem}
\usepackage{float}
\usepackage{array,booktabs}

\usepackage[numbers,sort&compress]{natbib}

\usepackage[dvipsnames]{xcolor}
\usepackage{tikz}
\usepackage{tikz-cd}
\usepackage{tensor}

\usepackage[utf8]{inputenc} 
\definecolor{azure}{rgb}{0.0, 0.5, 1.0}
\definecolor{darkblue}{rgb}{0.15,0.35,0.7}
\definecolor{reddish}{rgb}{0.65, 0.2, 0.2}
\definecolor{brandeisblue}{rgb}{0.0, 0.44, 1.0}
\definecolor{ceruleanblue}{rgb}{0.16, 0.32, 0.75}
\definecolor{indigo(dye)}{rgb}{0.0, 0.25, 0.42}
\usepackage[linktocpage=true]{hyperref}
\hypersetup{
colorlinks=true,
citecolor=ceruleanblue,
linkcolor=ceruleanblue,
urlcolor=ceruleanblue,
pdfauthor={},
pdftitle={},
pdfsubject={}
}

\newcommand{\overbar}[1]{\mkern 1.5mu\overline{\mkern-1.5mu#1\mkern-1.5mu}\mkern 1.5mu}

\newcommand{\TTbar}{T\overbar{T}}

\DeclareMathOperator{\tr}{\text{tr}}

\DeclareFontEncoding{LS1}{}{}
\DeclareFontSubstitution{LS1}{stix}{m}{n}
\DeclareSymbolFont{stixsymbols}{LS1}{stixscr}{m}{n}
\SetSymbolFont{stixsymbols}{bold}{LS1}{stixscr}{b}{n}
\DeclareMathSymbol{\kay}{\mathalpha}{stixsymbols}{"6B}
\DeclareMathSymbol{\hay}{\mathalpha}{stixsymbols}{"68}

\makeatletter
\renewcommand\section{\@startsection {section}{1}{\z@}%
                               {-3.5ex \@plus -1ex \@minus -.2ex}
                               {2.3ex \@plus.2ex}%
                               {\normalfont\large\bfseries}}
\renewcommand\subsection{\@startsection{subsection}{2}{\z@}%
                                 {-3.25ex\@plus -1ex \@minus -.2ex}%
                                 {1.5ex \@plus .2ex}%
                                 {\normalfont\bfseries}}
\makeatother

\makeatletter
\newcommand*\bigcdot{\mathpalette\bigcdot@{.5}}
\newcommand*\bigcdot@[2]{\mathbin{\vcenter{\hbox{\scalebox{#2}{$\m@th#1\bullet$}}}}}
\makeatother
\newcommand{\deq}{\stackrel{\bigcdot}{=}}

\newfont{\goth}{ygoth.tfm scaled 1200}                   

\numberwithin{equation}{section}

\def\a{\alpha}
\def\b{\beta}

\newcommand{\bea}{\begin{eqnarray}}
\newcommand{\eea}{\end{eqnarray}}
\newcommand{\ba}{\begin{array}}
\newcommand{\ea}{\end{array}}

\def\double #1{#1{\hbox{\kern-2pt $#1$}}}

\newcommand{\bsubeq}{\begin{subequations}}
\newcommand{\esubeq}{\end{subequations}}

\def\tr{{\rm tr}}

\renewcommand\baselinestretch{1.25}

\allowdisplaybreaks

\begin{document}
\begin{titlepage}
\begin{flushright}
\today
\end{flushright}
\vspace{5mm}

\begin{center}
{\Large \bf Integrable auxiliary field sigma models and \\ the metric approach to $T\bar{T}$-like deformations}
\end{center}

\begin{center}

{\bf
Nicola Baglioni${}^{a}$,
Daniele Bielli${}^{b,c}$,
Parita Shah${}^{b}$,\\
Gabriele Tartaglino-Mazzucchelli${}^{a}$
}
\\
\vspace{5mm}

\footnotesize{
${}^{a}$
{\it 
School of Mathematics and Physics, University of Queensland,
\\
 St Lucia, Brisbane, Queensland 4072, Australia}
}
\\~\\
\footnotesize{
${}^{b}$
{\it 
Department of Mathematical Sciences,
\\Durham University, Durham DH1 3LE, UK}
}
\\~\\
\footnotesize{
${}^{c}$
{\it 
Asia Pacific Center for Theoretical Physics (APCTP), \\ 
Postech, Pohang 37673, 
Korea}
}
\vspace{2mm}
~\\
\texttt{
n.baglioni@uq.edu.au,
d.bielli4@gmail.com,
parita.r.shah@durham.ac.uk,
g.tartaglino-mazzucchelli@uq.edu.au}
\\
\vspace{2mm}

\end{center}

\begin{abstract}
\baselineskip=14pt
\noindent
We describe a family of two-dimensional metric deformations that preserve integrability. The seeds are (classically) conformal integrable field theories and the deformations are directly related to different classes of auxiliary field sigma models. We introduce a master action connecting deformations by Lie-algebra-valued auxiliary fields, the so-called $\nu$-frame, to deformations by two Lorentz-charged scalars. These auxiliary scalar fields can be identified as Beltrami differentials of an auxiliary two-dimensional metric, providing a direct connection with the metric approach to $T\bar{T}$-like deformations. For Weyl-invariant seed matter theories, the auxiliary metric sector is described by a reduced gravitational potential. We show that the resulting framework preserves classical Lax integrability whenever the seed matter theory admits a Lax representation of its equations of motion on a generic curved two-dimensional spacetime. Finally, within the Lorentz-invariant family and the zweibein deformations of conformal matter theories considered here, we identify $\TTbar$ as the only deformation that can be interpreted as a field-dependent coordinate transformation.

\end{abstract}
\vspace{5mm}

\vfill
\end{titlepage}

\renewcommand{\thefootnote}{\arabic{footnote}}
\setcounter{footnote}{0}

\tableofcontents{}
\vspace{0.5cm}
\bigskip\hrule

\section{Introduction}
\label{sec:introduction}

The $T\bar T$ deformation provides a particularly useful example of how an irrelevant interaction can modify a quantum field theory while retaining exact control over part of its dynamics. Defined by a composite operator quadratic in the stress-energy tensor, it admits closed flow equations for finite-volume energy levels as well as other observables, independently of whether the undeformed theory is integrable~\cite{Zamolodchikov:2004ce,Smirnov:2016lqw,Cavaglia:2016oda}. For integrable seed theories, the deformation also preserves integrability and has a simple interpretation in terms of a universal dressing of the factorised scattering matrix~\cite{Smirnov:2016lqw}. These properties, together with connections to effective strings and holography~\cite{Cavaglia:2016oda,McGough:2016lol}, have motivated the study of more general stress-tensor flows. Among them, the classically marginal root-$T\bar T$ deformation supplies another distinguished integrability preserving deformation for broad classes of two-dimensional sigma models~\cite{Rodriguez:2021tcz,Babaei-Aghbolagh:2022uij,Ferko:2022cix,Babaei-Aghbolagh:2022leo,Borsato:2022tmu}. See \cite{Jiang:2019epa,Ferko:2021loo,He:2025ppz} for some pedagogical reviews on $\TTbar$-like deformations.

At the Lagrangian level, solving such flows directly can require studying complicated nonlinear differential equations. Geometric and auxiliary field formulations offer an alternative in that the deformed theory is obtained by solving algebraic equations for additional fields. An early geometric interpretation of $T\bar T$ identifies its classical equations with those of the seed theory written in field-dependent coordinates~\cite{Conti:2018tca,Conti:2019dxg}. Related descriptions in terms of topological gravity and random geometry~\cite{Dubovsky:2017cnj,Cardy:2018sdv,Dubovsky:2018bmo} connect this observation to gravitational dressing. In particular, in~\cite{Tolley:2019nmm}, $T\bar T$ was formulated as a coupling to two-dimensional massive gravity involving an auxiliary zweibein and a reference zweibein. More general relative-vielbein potentials accommodate wider families of stress-tensor deformations, including combined $T\bar T$ and root-$T\bar T$ flows~\cite{Babaei-Aghbolagh:2024hti,Tsolakidis:2024wut}.  Metric methods have also proved useful beyond two dimensions, for example, for nonlinear electrodynamics~\cite{Conti:2022egv,Ferko:2024yhc} and non-Abelian Yang--Mills flows~\cite{Ferko:2024yua}.

A different auxiliary field construction arose from four-dimensional duality-invariant nonlinear electrodynamics. Prominent examples are Born--Infeld electrodynamics and its supersymmetric extensions~\cite{cecotti:1987,Bagger:1996wp,Kuzenko:2000tg,Kuzenko:2000uh,Tseytlin:1999dj}, as well as ModMax and its Born--Infeld-like and supersymmetric extensions~\cite{Bandos:2020jsw,Kuzenko:2021cvx,Bandos:2021rqy}. These theories are related by stress-tensor and supercurrent-squared flows~\cite{Conti:2018jho,Ferko:2019oyv,Ferko:2022iru}, and have also been used to study the preservation of vanishing birefringence~\cite{Ferko:2023ruw}.
In the Ivanov--Zupnik approach, electromagnetic duality can be implemented as a symmetry of the auxiliary field interaction~\cite{Ivanov:2001ec,Ivanov:2002ab,Ivanov:2003uj,Kuzenko:2013gr}. This makes the use of auxiliary field formalisms particularly effective for engineering duality-invariant stress-tensor flows~\cite{Ferko:2023wyi}. Motivated by these auxiliary field techniques and by integrable stress-tensor deformations of sigma models~\cite{Borsato:2022tmu}, auxiliary field sigma models (AFSMs) were introduced, in which the principal chiral model (PCM) is coupled to Lie-algebra-valued fields $v_\pm$ with algebraic equations of motion~\cite{Ferko:2024ali}. An arbitrary function $E$ of the product $\operatorname{tr}(v_+^2)\operatorname{tr}(v_-^2)$ parametrises an infinite family of stress-tensor deformations including $T\bar{T}$ and root-$T\bar{T}$. Crucially, Lax integrability and the associated Maillet structure of all theories in the family can be established by using the algebraic auxiliary EOM, without solving them explicitly ~\cite{Ferko:2024ali,Bielli:2024ach}.

The two-dimensional AFSM construction extends to PCMs with a Wess--Zumino term~\cite{Bielli:2024ach}, non-Abelian T-dual models~\cite{Bielli:2024khq}, symmetric and semi-symmetric spaces, as well as more general $\mathbb Z_N$ cosets~\cite{Cesaro:2024ipq,Bielli:2024oif}, (bi-)Yang--Baxter deformations~\cite{Bielli:2024fnp} and integrable sectors of gravity \cite{Cesaro:2025msv,Bielli:2026vit}. Recent works also provided a common description of the non-local charges, Yangian symmetry and Maillet brackets arising in such theories~\cite{Bielli:2026ggs}, and a further extension to the $\lambda$-model~\cite{Ferko:2026Lambda}. A four-dimensional Chern--Simons realisation of the AFSM, including its Wess--Zumino extension, was also constructed in~\cite{Fukushima:2024nxm}. Moreover, allowing auxiliary field interactions involving higher-degree chiral invariants preserves a universal Lax structure~\cite{Bielli:2024ach}. The auxiliary description thereby supports the construction of higher-spin currents and the systematic integration of Smirnov--Zamolodchikov flows~\cite{Smirnov:2016lqw}, including genuinely higher-spin deformations not exhausted by functions of the stress tensor~\cite{Bielli:2025uiv}.

The relation between different auxiliary field formulations has recently been clarified in~\cite{Baglioni:2025tsc}. In four-dimensional electrodynamics, the scalar $\mu$-frame representation of Ivanov and Zupnik \cite{Ivanov:2003uj} is related, through Legendre transformations and field redefinitions, to the Russo--Townsend formulation~\cite{Russo:2025Simplified} and its Courant--Hilbert description. Analogous structures organise two-dimensional integrable sigma models~\cite{Babaei-Aghbolagh:2025uoz,BabaeiAghbolagh:2025CH}. In particular, a two-dimensional $\mu$-frame replaces the Lie-algebra-valued auxiliaries with a scalar, which simplifies the algebraic identities underlying the Lax connection, and retains the relevant Poisson structure~\cite{Baglioni:2025tsc}. These developments suggest that the geometric and the AFSM constructions may be different auxiliary descriptions of the same deformed theories. The central result of the paper is the precise description of this equivalence.

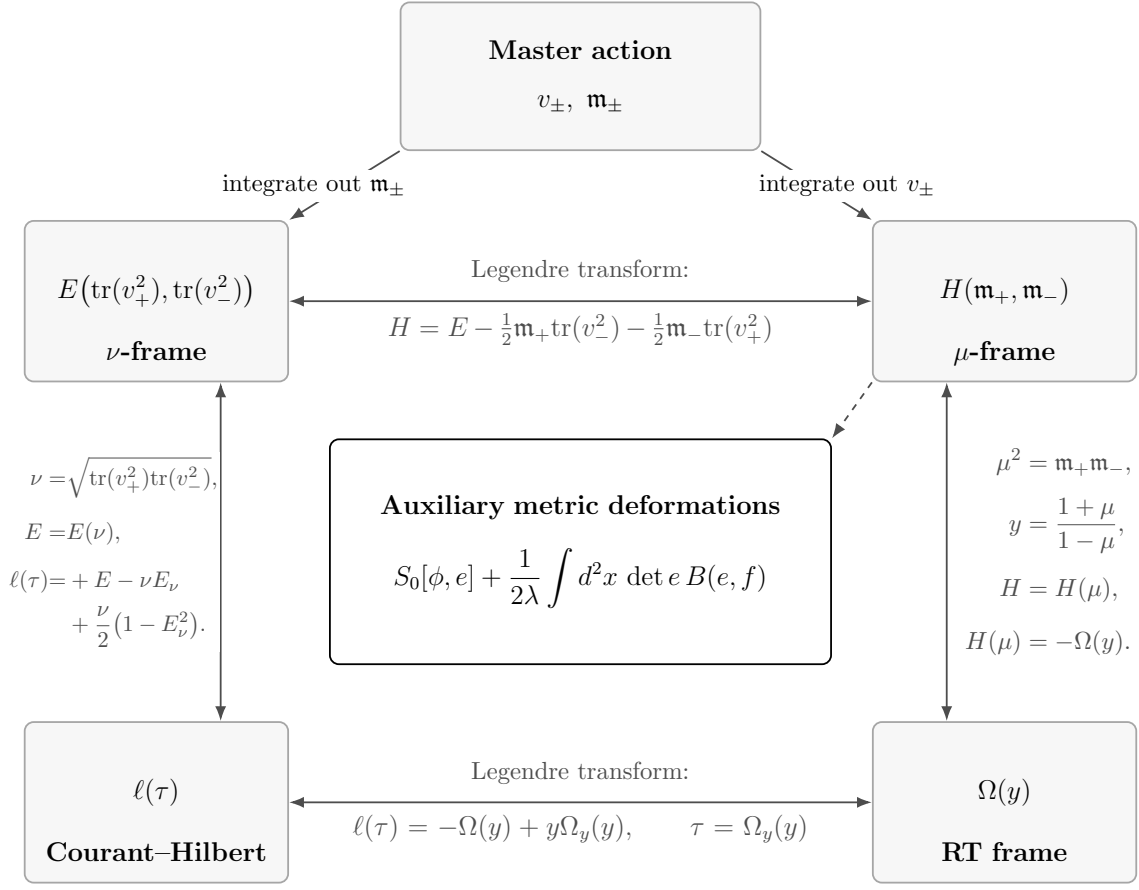
\begin{figure}[H]
    \centering
    \hspace*{-0.4cm}
    \scalebox{0.85}{%
    \begin{tikzpicture}[
        >=Latex,
        box/.style={
            draw=black!35,
            rounded corners=4pt,
            thick,
            fill=black!3,
            minimum width=4.1cm,
            minimum height=1.65cm,
            text width=3.7cm,
            align=center,
            inner sep=4pt
        },
        topbox/.style={
            draw=black!35,
            rounded corners=4pt,
            thick,
            fill=black!3,
            minimum width=5.6cm,
            minimum height=2.3cm,
            text width=5.0cm,
            align=center,
            inner sep=6pt
        },
        centerbox/.style={
            draw=black,
            rounded corners=4pt,
            thick,
            minimum width=7.8cm,
            minimum height=3.5cm,
            text width=7.1cm,
            align=center,
            inner sep=6pt
        },
        eqarr/.style={
            <->,
            thick,
            black!70
        },
        dirarr/.style={
            ->,
            thick,
            black!70
        },
        linkarr/.style={
            dashed,
            ->,
            thick,
            black!70
        },
        lab/.style={
            fill=white,
            inner sep=2pt,
            font=\small,
            align=center
        },
        labformula/.style={
            fill=white,
            inner sep=2pt,
            font=\small,
            align=center
        }
    ]


    \node[topbox] (master) at (0,7.4)
    {
        \textbf{Master action}
        \\[0.6ex]
        \(v_\pm,\ \mathfrak m_\pm\)
    };


    \node[box] (nu) at (-6.6,  3.9)
    {
        \[
        E\bigl(\mathrm{tr}(v_+^2),\mathrm{tr}(v_-^2)\bigr)
        \]
        \\[-0.5ex]
        \strut \textbf{\(\nu\)-frame}
    };

    \node[box] (mu) at ( 6.6,  3.9)
    {
        \[
        H(\mathfrak m_+,\mathfrak m_-)
        \]
        \\[-0.5ex]
        \strut \textbf{\(\mu\)-frame}
    };

    \node[box] (rt) at ( 6.6, -3.9)
    {
        \[
        \Omega(y)
        \]
        \\[-0.5ex]
        \strut \textbf{RT frame}
    };

    \node[box] (ch) at (-6.6, -3.9)
    {
        \[
        \ell(\tau)
        \]
        \\[-0.5ex]
        \strut \textbf{Courant--Hilbert}
    };


    \node[centerbox] (new) at (0,0)
    {
        \textbf{Auxiliary metric deformations}
        \[
        S_0[\phi,e]
        +\frac{1}{2\lambda}
        \int d^2x\,\det e\,B(e, f)
        \]
    };


    \draw[dirarr] (master.south west) -- (nu.north east);
    \draw[dirarr] (master.south east) -- (mu.north west);

    \node[lab] at (-4.15,5.70)
        {integrate out \(\mathfrak m_\pm\)};

    \node[lab] at (4.15,5.70)
        {integrate out \(v_\pm\)};


    \draw[eqarr]
        (nu) -- (mu)
        node[lab,above,pos=.5,yshift=6pt]
        {Legendre transform:}
        node[below,pos=.5,yshift=-1pt,text width=6.0cm,align=center]
        {$
        H
        =
        E
        -\frac12\mathfrak m_+\mathrm{tr}(v_-^2)
        -\frac12\mathfrak m_-\mathrm{tr}(v_+^2)
        $};


    \draw[eqarr]
        ([xshift=-0.90cm]mu.south)
        --
        ([xshift=-0.90cm]rt.north)
        node[
            labformula,
            right,
            pos=.5,
            xshift=0pt,
            text width=3.0cm
        ]
        {$\displaystyle
        \begin{aligned}
        \mu^2 &= \mathfrak m_+\mathfrak m_-,
        \\[0.7ex]
        y &= \frac{1+\mu}{1-\mu},
        \\[0.7ex]
        H &= H(\mu),
        \\[0.7ex]
        H(\mu) &= -\Omega(y).
        \end{aligned}
        $};


    \draw[eqarr]
        (ch) -- (rt)
        node[lab,above,pos=.5,yshift=6pt]
        {Legendre transform:}
        node[below,pos=.5,yshift=-1pt,text width=7.4cm,align=center]
        {$
        \ell(\tau)
        =
        -\Omega(y)+y\Omega_y(y),
        \qquad
        \tau=\Omega_y(y)
        $};


    \draw[eqarr]
        ([xshift=1cm]nu.south)
        --
        ([xshift=1cm]ch.north)
        node[
            labformula,
            left,
            pos=.5,
            xshift=0pt,
            text width=3.2cm,
            font=\footnotesize
        ]
        {$\displaystyle
        \begin{aligned}
        \hspace{-1mm}\nu
        =&
        \sqrt{
            \mathrm{tr}(v_+^2)
            \mathrm{tr}(v_-^2)
        },
        \\[0.8ex]
        \hspace{-1mm}E =& E(\nu),
        \\[0.8ex]
        \ell(\tau)
        \hspace{-1mm}=&
        +E-\nu E_\nu
        \\
        \hspace{-1mm} &
        +\frac{\nu}{2}
        \bigl(1-E_\nu^2\bigr).
        \end{aligned}
        $};


\draw[linkarr] (mu.south west) -- (new.north east);
    \end{tikzpicture}
    }

    \caption{
        Relations between the \(\nu\)-frame, \(\mu\)-frame,
        the Russo--Townsend auxiliary formulation,
        the Courant--Hilbert description,
        the auxiliary metric deformation action,
        and the master action.
    }
    \label{fig:frames-map}
\end{figure}

 In section \ref{sec_1} we introduce a master Lagrangian containing both $v_\pm\in \mathfrak{g}$ and two scalar auxiliaries, $\mu_{2+}$ and $\mu_{2-}$, respectively conjugate to $\mathrm{tr}(v_-^2)$ and $\mathrm{tr}(v_+^2)$. Eliminating the scalars recovers the $\nu$-frame descriptions of AFSM, while eliminating $v_\pm$ leads to a frame characterised by the two charged scalar fields, with $\mathfrak m_\pm=2\mu_{2\pm}$ and interaction $H(\mathfrak m_+,\mathfrak m_-)$. Unlike the single-scalar formulation, the so-called $\mu$-frame, this description keeps the two Lorentz-charged variables independent and also accommodates Lorentz-breaking interactions. The Lorentz-invariant sector is recovered when $H$ depends only on the product $\mathfrak m_+\mathfrak m_-$, reducing back to the $\mu$-frame description of \cite{Baglioni:2025tsc}. In this setting, the $E$ and $H$ functions are Legendre transforms of each other. The corresponding auxiliary field descriptions and their relations are summarised in Figure~\ref{fig:frames-map}.

We focus on the geometric interpretation in section \ref{sec_3}.
The geometric meaning of the charged frame follows from the auxiliary zweibein one-forms
\begin{equation}
 e^+=\Omega\bigl(d\sigma^++\mathfrak m_-\,d\sigma^-\bigr),
 \qquad
 e^-=\Omega\bigl(d\sigma^-+\mathfrak m_+\,d\sigma^+\bigr).
 \label{eq:intro-beltrami}
\end{equation}
The fields $\mathfrak m_\pm$ are precisely auxiliary Beltrami differentials parametrising the conformal metric.

 For classically Weyl-invariant matter, the conformal factor $\Omega$ does not enter the seed theory action, and its algebraic elimination from the gravitational potential leads to the interaction function $H$. Thus, such a scalar interaction function has a direct interpretation as a reduced auxiliary metric potential. In this work, we use the PCM as our main example; however, the same results can be extended to the common sigma-model structure described in Appendix~\ref{app_A0}. Indeed, the key result of this work is that this metric approach should be applicable to any model that is integrable on any curved worldsheet.

The PCM admits a Lax formulation on a general worldsheet metric~\cite{Kluson:2007vw, Driezen:2021cpd}. A closely related geometric perspective was developed in~\cite{Sakamoto:2025hwi}. Starting from four-dimensional Chern--Simons theory, they constructed curved-background actions and Lax connections for a class of integrable degenerate $\mathcal E$-models, using Beltrami variables, and then specialised to the metrics associated with $T\bar T$ and root-$T\bar T$. Our work generalises this approach to any field-dependent metric that is obtained by integrating out the auxiliary metric. This distinction is key to our analysis.

In section \ref{integrability} we show that a seed theory admitting a Lax representation of its equations of motion on a generic two-dimensional curved spacetime continues to admit a Lax representation after coupling it to an algebraic auxiliary metric with a matter-independent potential. The equations of motion of the physical fields in the reduced theory coincide with the seed equations evaluated on the auxiliary metric solution, with the metric held fixed in the variation of the seed action. The corresponding curved Lax connection may therefore be evaluated on the same auxiliary solution, providing a sufficient criterion for weak classical integrability for arbitrary admissible $H$, rather than only for the distinguished $T\bar T$ and root-$T\bar T$ deformations considered in \cite{Sakamoto:2025hwi}. 

Mutual Poisson commutativity of the spectral invariants of the monodromy matrix, which leads to strong integrability, requires additional Hamiltonian information. For models equivalent to the previously studied AFSM formulations, this stronger statement follows from the known Maillet structure~\cite{Ferko:2024ali,Bielli:2024ach,Baglioni:2025tsc,Bielli:2026ggs}.

Our analysis also provides an appropriate Lagrangian extension of the geometric interpretation developed in~\cite{Ferko:2025bhv}, based on the earlier field-dependent-metric description of the equations of motion of four-dimensional duality-invariant electrodynamics of~\cite{Ferko:2024yhc}. 
For a specific class of symmetric-space sigma models, which includes the PCM as an example, it was shown that the deformed equations of motion could be described as undeformed equations on a field-dependent metric $h$ with $\det h=\det\eta=-1$. What was missing was an appropriate action principle extension of this result, which is presented in this work.

Finally, we stress that in our approach, a field-dependent metric need not arise from a field-dependent coordinate transformation. We investigate this distinction in section \ref{sec_coc} by imposing coordinate-integrability conditions on the Beltrami coframe. Within the Lorentz-invariant family $H=H(\tilde\mu)$, $\tilde\mu=\mathfrak m_+\mathfrak m_-$ the coordinate-integrability conditions impose $H=k_1/(1-\tilde\mu)+k_2$, namely the $T\bar T$ potential up to normalisation and an additive constant. This singles out $T\bar T$ as the only field-dependent coordinate transformation, excluding a coordinate transformation interpretation for the larger family of integrability-preserving metric deformations. 

We leave some of the longer calculations for the appendices. In \ref{app_A0} we extend the arguments from the PCM to other seed theories, including Yang-Baxter and T-dual models. In \ref{app_A} we compute the expression of the deformed stress-energy tensor and in \ref{app_2} we report some details of the calculation carried out in section \ref{sec_coc}.

\medskip
\noindent\textit{{\bf Note added.}} While this work was being completed, the preprint~\cite{Fukushima:2026CHgravity} appeared, presenting a gravitational realisation of Courant--Hilbert deformations through a spectral decomposition of the relative zweibein. Its results overlap with our identification of scalar auxiliary formulations with gravitational ones. The two analyses are, however, complementary. In particular, our paper aims to emphasise the charged AFSM--Beltrami dictionary and its relation with the $\mu$-frame of \cite{Baglioni:2025tsc}, the conditions for preservation of classical Lax integrability, and the distinction between metric deformations and field-dependent coordinate transformations.

\section{Integrable deformations as coupling to scalar auxiliary fields}\label{sec_1}
Consider the family of integrable deformations of the Principal Chiral Model (PCM) first introduced in \cite{Ferko:2024ali}, also known as the auxiliary field sigma model (AFSM):\footnote{We will use the light-cone conventions adopted in \cite{Ferko:2024ali}:
\begin{equation}\label{lc_conventions}
\sigma^\pm=\frac12(\tau\pm \sigma)\,, \qquad \eta_{+ -}=\eta_{-+}=-2 \,,
\qquad 
\eta^{+ -}=\eta^{-+}=-\frac{1}{2} \,.
\end{equation}}
\begin{equation}\label{L_AFSM}
\begin{aligned}
\mathcal{L}_{\text{AFSM}}
={}&
\frac{1}{2}\operatorname{tr}(j_+j_-)
+\operatorname{tr}
\left(
j_-v_+ + j_+v_- + v_+v_-
\right)
+E(v_+,v_-)\,.
\end{aligned}
\end{equation}
The physical field of the theory is the group element $g(\tau,\sigma)\in G$, encoded in the Lie-algebra-valued Maurer-Cartan current $j_\pm:=g^{-1}\partial_{\pm}g \in \mathfrak{g}$, while the fields $v_\pm \in \mathfrak{g}$ are auxiliary, in the sense that their equations of motion are purely algebraic. These equations explicitly depend on the interaction function $E(v_+,v_-)$, whose detailed structure selects an element in the infinite family of theories \eqref{L_AFSM}. Given a choice of interaction function, explicitly solving the auxiliary field equations of motion (EOM) for $v_{\pm}$, and substituting back into the action, leads to a deformed theory fully expressed in terms of the physical fields. The seed theory in the family, namely the PCM, corresponds to the choice $E=0$, in which case the auxiliary field EOM reduces to $v_{\pm}=-j_{\pm}$ such that \eqref{L_AFSM} becomes 
\begin{equation}
\mathcal{L}_{\text{PCM}}=-\frac12 \mathrm{tr}(j_+ j_-)\, .
\end{equation}
The family of theories \eqref{L_AFSM} has been extensively studied in recent years for various reasons. Firstly, the model has been proven to be (classically) integrable for any $E$ that does not depend on ``mixed traces'' of $v_+$ and $v_-$, see, e.g., \cite{Bielli:2024ach}. That is, $E$ can depend on traces of the form $\mathrm{tr}(v_\pm^2),\, \mathrm{tr}(v_\pm^3),\, \mathrm{tr}(v_\pm^4)\,,\dots $ and the (physical) EOMs will be equivalent to the flatness of a Lax connection. Notice that this is only a sufficient condition.\footnote{For example, assuming that $E$ depends also on $p=\mathrm{tr}(v_+v_-)$ does not seem to necessarily break the integrability of the model, although it could produce constraints on the possible form of the interaction function $E$, see \cite{Baglioni:2025tsc}.}
For such integrable deformations, it was observed that when $E$ is restricted to be a function of $\mathrm{tr}(v_+^2)\mathrm{tr}(v_-^2)$, the model corresponds to all possible stress-tensor deformations of the PCM \cite{Ferko:2024ali}. This is an infinite family of integrable deformations that also includes the celebrated $T\bar T$- and $\sqrt{T\bar T}$-deformations. At the Lagrangian level, these deformations correspond to solving the differential equations
\begin{equation}
    \frac{\partial \mathcal L}{\partial \lambda}=\mathcal O_\lambda\,,
\end{equation}
where $\mathcal O_\lambda$ is some function of the stress-energy tensor.
Specifically the $T\bar T$ and the root-$T\bar T$ operators are:
\begin{equation}
\begin{aligned}
\mathcal{O}_{T\bar T}=&~T_{\mu\nu}T^{\mu\nu}-(T^{\mu}_{\,\,\,\mu})^2\,,
\\
\mathcal{R}_{\sqrt{T\bar T}}=&~\frac{1}{\sqrt{2}}\sqrt{T_{\mu\nu}T^{\mu\nu}-\frac{1}{2}(T^{\mu}_{\,\,\,\mu})^2}\,.
\end{aligned}
\end{equation}
Furthermore, the model \eqref{L_AFSM} opens the possibility of studying higher-spin deformations, corresponding to traces of more than two $v_\pm$ (e.g., $\mathrm{tr}(v_\pm^3)$) in $E$. The structure of these higher-spin deformations remains largely unexplored, and a systematic study was initiated in \cite{Bielli:2025uiv} by using auxiliary fields. Finally, let us stress that these results have been non-trivially extended to a broad range of integrable models that have a structure that is similar to that of the PCM. In particular, this construction works for coset models \cite{Bielli:2024oif}, T-dual models \cite{Bielli:2024khq}, (bi)-Yang-Baxter deformed models \cite{Bielli:2024fnp}, WZW models \cite{Bielli:2024ach}. In the remainder of this article, we will focus on the deformed PCM to simplify the discussion. However, all of the results we will mention hold for this much broader class of models.

Let us start from \eqref{L_AFSM}, assuming $E$ is allowed to depend on both $\mathrm{tr}(v_+^2)$ and $\mathrm{tr}(v_-^2)$, leading to a non-relativistic model in general. The goal of this section is to introduce a new set of scalar auxiliary fields that still describes the same models as \eqref{L_AFSM}. The idea is similar to what was already done in \cite{Baglioni:2025tsc}, where we traded the AFSM with  $E\Bigl(\mathrm{tr}(v_+^2)\mathrm{tr}(v_-^2)\Bigl)$ for a model with a single scalar auxiliary field $\mu$ (so-called $\mu$-frame).
Let us present a generalization of that procedure.
Consider the following master Lagrangian
\begin{equation}\label{L_master}
\mathcal{L}_{\mathrm M}
=
\frac{1}{2}\operatorname{tr}(j_+j_-)
\!+\!\operatorname{tr}
\left(
j_-v_+ \!+\! j_+v_- \!+\! v_+v_-
\right)\!+\!\mu_{2-}\operatorname{tr}(v_+^2)
\!+\!\mu_{2+}\operatorname{tr}(v_-^2)
\!+\!H(\mu_{2+},\mu_{2-}) \, ,
\end{equation}
where we have introduced two additional scalar auxiliary fields $\mu_{2\pm}$, with the labels $2\pm$ indicating their Lorentz weight, and $H$ is again an arbitrary interaction function. The fields $v_\pm$ now couple to the physical fields as well as to the new auxiliary fields $\mu_{2\pm}$, which in addition self-interact through $H$. Let us now consider the equations of motion of $\mu_{2\pm}$ first, which are given by
\begin{equation}
    \mathrm{tr}(v_+^2)=-H_{\mu_{2-}}\,,\quad \mathrm{tr}(v_-^2)=-H_{\mu_{2+}}\,,
\end{equation}
where we use subscripts to indicate partial derivatives of the function $H$ with respect to $\mu_{2\pm}$. Taking these two equations as the definition of the conjugate Legendre momenta, we see that, after integrating out $\mu_{2\pm}$, the action reduces to \eqref{L_AFSM} when
\begin{equation}
    E= H+\mathrm{tr}(v_+^2) \mu_{2 -}+\mathrm{tr}(v_-^2) \mu_{2 +}\,.
\end{equation}
This means that choosing a function $H$ and integrating out $\mu_{2\pm}$ first gives us the $\nu$-frame AFSM with interaction function $E$ equal to the double Legendre transform of $H$. This relation constitutes the top side of the square depicted in Figure~\ref{fig:frames-map}. 
For the rest of our discussion, it is convenient to rescale the fields $\mu_{2\pm}$ as
\begin{equation}
\mathfrak m_- := 2\mu_{2-},
\qquad
\mathfrak m_+ := 2\mu_{2+},
\qquad
\tilde{\mu} := \mathfrak m_+\mathfrak m_- = 4\mu_{2-}\,\mu_{2+}.
\end{equation}
Following the conventions where $j_+$ has Lorentz weight $+1$ and $j_-$ has
weight $-1$, the scalar auxiliary fields have the following weights,
\begin{equation}
\operatorname{wt}(\mathfrak m_-)=-2,
\qquad
\operatorname{wt}(\mathfrak m_+)=+2,
\qquad
\operatorname{wt}(\tilde{\mu})=0.
\end{equation}

Starting from the master Lagrangian \eqref{L_master}, we can consider integrating out the $v_\pm$ auxiliary fields first instead. Therefore, the equations of motion for $v_\pm$ are given by
\begin{equation}
j_-+v_-+\mathfrak m_-v_+=0,
\qquad
j_++v_++\mathfrak m_+v_-=0.
\end{equation}
For $\tilde \mu\neq1$, their solution is
\begin{equation}\label{mPCM-v-EOM}
v_+
=
-\frac{j_+-\mathfrak m_+j_-}{1-\tilde \mu},
\qquad
v_-
=
-\frac{j_--\mathfrak m_-j_+}{1-\tilde \mu}.
\end{equation}
Substituting the above into \eqref{L_master} gives us a model where the deformation is equivalent to a coupling to the scalar auxiliary fields $\mathfrak m_{\pm}$, which we shall refer to as the ``$\frak m_\pm$-frame'',
\begin{equation}\label{mpm_frame}
\mathcal{L}_{\mathfrak m}
=
\frac{
\mathfrak m_+\mathrm{tr}(j_-^2)+\mathfrak m_-\mathrm{tr}(j_+^2)-(1+\mathfrak m_-\mathfrak m_+)\mathrm{tr}(j_-j_+)
}{
2(1-\mathfrak m_-\mathfrak m_+)
}
+
H(\mathfrak m_+,\mathfrak m_-).
\end{equation}
This auxiliary model with two scalar fields $\frak m_\pm$ is therefore a generalization of what we called the $\mu$-frame in \cite{Baglioni:2025tsc}. The $\mu$-frame is obtained when $H$ only depends on $\frak m_+ \frak m_-$, which means that the results of \cite{Baglioni:2025tsc} can be recovered from the generic $\frak m_\pm$ case by setting\footnote{Also note that, in the conventions of \cite{Baglioni:2025tsc}, where only one scalar auxiliary field $\mu$ was used,  $\tilde{\mu}=\mathfrak m_+\mathfrak m_-$ can simply be written as $\mu^2$.} 
\begin{equation}\label{mpm to mu}
    \frak m_\pm =\mu \sqrt{\frac{\mathrm{tr}(j_\pm^2)}{\mathrm{tr}(j_\mp^2)}}\,.
\end{equation}

\paragraph{Relation with Courant-Hilbert formulation.}
Let us stress that the models considered above are also equivalent to the so-called Courant-Hilbert (CH) formulation \cite{Babaei-Aghbolagh:2025cni,Fukushima:2024nxm,Fukushima:2026gan,Fukushima:2025tlj,Fukushima:2026CHgravity}. The CH formulation characterizes the deformed theory in terms of an auxiliary variable subject to a partial differential equation, which arises as the integrability condition of the deformed equations of motion. As shown in \cite{Baglioni:2025tsc}, when restricted to relativistic deformations, this formulation is equivalent to the auxiliary field approach discussed above.
The relativistic restriction amounts to taking the interaction functions to depend only on the Lorentz-invariant combinations $E=E(\nu)$, with $\nu=\sqrt{\mathrm{tr}(v_+^2)\mathrm{tr}(v_-^2)}$ and, equivalently, $H=H(\mu)$. The CH construction admits two Legendre-dual descriptions, in terms of an interaction function $\ell(\tau)$ or a function $\Omega(y)$. In the first description, the CH Lagrangian is,
\begin{equation}\label{PDE_intro}
\mathcal L= \ell(\tau)-\frac{2 U}{\dot\ell(\tau)} \,\, ,
\qquad \text{with} \qquad 
\tau=V+\frac{U}{\dot\ell(\tau)^2} \,\, ,
\end{equation}
and, for the PCM, if we define 
\begin{equation}
    S = - \frac{1}{2} \tr \left( j_+ j_- \right)\,,\quad P^2 = \frac{1}{4} \left( \tr \left( j_+ j_+ \right) \tr \left( j_- j_- \right) - \left( \tr \left( j_+ j_- \right) \right)^2 \right)\,,
\end{equation}
and 
\begin{equation}\label{U-V-def}
U=\frac{1}{2}\left(\sqrt{S^2+P^2}-S\right) \,\, ,
\qquad \qquad 
V=\frac{1}{2}\left(\sqrt{S^2+P^2}+S\right)\,\, .
\end{equation}
Thus, given $\ell(\tau)$, the second equation in \eqref{PDE_intro} determines the auxiliary variable $\tau$ in terms of the physical Lorentz scalars $U$ and $V$, $\tau=\tau(U,V)$. The Legendre-dual description is defined by
\begin{equation}\label{Legendre_Omega_l}
\ell(\tau)=-\Omega(y)+y\Omega_y(y),\qquad
\dot{\ell}=y,\qquad
\tau=\Omega_y.
\end{equation}
In terms of these quantities, the Lagrangian is
\begin{equation}\label{L_omega}
    \mathcal{L}=y\,V-\frac U y -\Omega(y)\,,
\end{equation}
and the second relation in \eqref{PDE_intro} is precisely the EOM of $y$, when $y$ is treated as an auxiliary field, making the equivalence with the auxiliary field formulation manifest. One of the other results of \cite{Baglioni:2025tsc} was the identification of these models with the $\nu$- and $\mu$-frame. The complete relation between these equivalent descriptions is summarized in Figure~\ref{fig:frames-map}. The corresponding interaction functions are related by \cite{Baglioni:2025tsc}
\begin{equation}\label{map_EHlOmega}
\begin{split}
    &H(\mu)=-\Omega(y)\,,\quad\quad \quad\qquad \quad \quad\quad y=\frac{1+\mu}{1-\mu}\,,\\[0.5em]
    &\ell(\tau)=\,\,\frac{\nu}{2}(1-E_\nu^2)-\nu E_\nu+E\,,\quad \,\,\,\,
    \tau=\frac{\nu}{2}(1-E_\nu)^2\,.
\end{split}
\end{equation}
These are therefore four descriptions of the same family of theories\footnote{Notice that, due to the property of the Legendre transformation, some models that are described by the $\nu$-frame cannot be described in the $\mu$-frame. One such example is the pure $\sqrt{T\bar T}$-flow.} and the above relation \eqref{map_EHlOmega} gives the mapping between the interaction functions in each of these frames.

The master Lagrangian \eqref{L_master} makes the connection between the so-called $\nu$-frame (where we have auxiliary fields $v_\pm$) and the $\frak m_\pm$-frame (where we only have scalar auxiliary fields) immediate. Compared to the more involved analysis of \cite{Baglioni:2025tsc}, it is now possible to go from one frame to the other by simply integrating out either set of auxiliary fields at the level of the master Lagrangian. The Lagrangian \eqref{mpm_frame} is interesting for several reasons. Most importantly, in this work we will show that the $\frak m_\pm$-frame model \eqref{mpm_frame} is equivalent to coupling the seed theory to an auxiliary metric. This gravitational approach was already introduced in the literature to study some specific stress tensor deformations; see \cite{Tolley:2019nmm,Dubovsky:2017cnj,Babaei-Aghbolagh:2024hti}. As we will show in the next sections, the Lagrangian \eqref{mpm_frame} naturally extends this approach to the entire family of stress-tensor deformations. Before diving into this, let us conclude this initial section with a few comments.  As mentioned earlier, the $\frak m_\pm$-frame can be extended to a broad range of models where the analogue of \eqref{mpm_frame} is
\begin{equation}\label{mpm_frame_general}
\mathcal{L}_{\mathfrak m}
=
\frac{
(1+\mathfrak m_-\mathfrak m_+)\mathcal{L}_{G}^{(0)}-\mathfrak m_+T^{(0)}_{--}-\mathfrak m_-T^{(0)}_{++}
}{
1-\mathfrak m_-\mathfrak m_+
}
+\mathcal{L}^{(0)}_{B}+
H(\mathfrak m_+,\mathfrak m_-).
\end{equation}
 In Appendix \ref{app_A0} we use a unified formalism to show the validity of \eqref{mpm_frame_general} for coset models, Yang-Baxter deformed models, and T-dual models. However, we do expect \eqref{mpm_frame_general} to hold even more generally.
In this general construction of the Lagrangian, the fields $\frak m_\pm$ couple to the undeformed components of the stress-energy tensor $T_{\pm\pm}^{(0)}$, and to the two-dimensional spacetime metric part of the non-linear sigma-model, $\mathcal L_G^{(0)}$, while the $B$-field contribution $\mathcal L_B^{(0)}$ is left untouched. 
This is the general structure that we have observed for all the models where these stress-tensor deformations preserve integrability. However, a crucial property of these seed theories is that they are classically scale invariant. In particular, if the stress-energy tensor has a non-vanishing trace, the coupling in \eqref{mpm_frame_general} would not be sufficient to capture any possible stress-tensor deformation. In fact, we will see in the next section that this observation mirrors the fact that a non-conformal seed theory will couple to the conformal factor of the auxiliary metric, spoiling this $\frak m_\pm$ picture. On the other hand, this indicates that a formulation similar to \eqref{mpm_frame} might apply to any CFT. This suggests that auxiliary metric deformations may preserve not only integrability, but also more general conformal structures.

We now comment on the integrability of this formulation. For the case of a single scalar field $\mu$, it was shown in \cite{Baglioni:2025tsc} that one can construct a Lax connection that is flat on shell. The corresponding Lax connection is obtained essentially from that in the $\nu$-frame by replacing the currents with their counterparts in the scalar frame. The Maillet form of the Poisson bracket was also shown to be preserved in the scalar frame. A similar analysis would produce the same conclusions in the case of two scalar fields $\frak m_{\pm}$ as well. Instead of repeating this analysis, in section \ref{integrability} we present a different approach to the integrability of these deformations, based on the metric approach developed in the next section. 

It must be noted that, although we focus on stress-tensor deformations here, in principle, a similar analysis can be carried out for higher-spin deformations as well. Instead of the terms $\mathrm{tr}(v_+^2)\mu_{2-}+\mathrm{tr}(v_-^2)\mu_{2+}$ in \eqref{L_master}, one can add a tower of scalars $\sum_{n=2}^m\mathrm{tr}(v_-^n)\mu_{n+}+\mathrm{tr}(v_+^n)\mu_{n-}$ to study spin-$m$ deformations. For this case, however, we do not expect the model to be equivalent to a metric coupling. The analysis is also more involved than in the spin-2 case, since the equations of motion for $v_\pm$ are no longer linear in the presence of these new couplings. Nevertheless, it should still be possible to introduce couplings to scalar auxiliary fields that describe higher-spin deformations of these models. We leave the exploration of this interesting direction to future work.

\section{Auxiliary metric deformations}\label{sec_3}
In this section, we make contact between the auxiliary field model we introduced in the previous section and the metric approach that has previously been developed in the literature. We will study deformations of the PCM in flat spacetime although, as we will argue later, this entire discussion will also hold for any fixed two-dimensional spacetime and for more general classes of models.

The main result of this section is that the action \eqref{mpm_frame_general} can be interpreted as coupling the seed theory to a metric.  
We showed in \eqref{mpm_frame} that the deformed Lagrangian reads
\begin{equation}\label{m_pm_metric}
 \mathcal L_\frak m= \frac{
\mathfrak m_+\mathrm{tr}(j_-^2)+\mathfrak m_-\mathrm{tr}(j_+^2)-(1+\mathfrak m_-\mathfrak m_+)\mathrm{tr}(j_-j_+)
}{2
(1-\mathfrak m_-\mathfrak m_+)
}
+
H(\mathfrak m_+,\mathfrak m_-)\,.
\end{equation}
This Lagrangian can be written as
\begin{equation}\label{Auxiliary_metric_Lag1}
\mathcal{L}_{\mathfrak m}
=
\frac{1}{4}\sqrt{-g}\,
g^{\alpha\beta}\operatorname{tr}(j_\alpha j_\beta)
+
H(\mathfrak m_+,\mathfrak m_-)
\end{equation}
as long as one defines the Weyl-invariant inverse metric density as
\begin{equation}\label{G^{AB}}
\mathcal{G}^{\alpha\beta}
:=
\sqrt{-g}\,g^{\alpha\beta}
=
\frac{1}{1-\tilde \mu}
\begin{pmatrix}
2\mathfrak m_- &-(1+\tilde \mu) \\
-(1+\tilde \mu) & 2\mathfrak m_+
\end{pmatrix},
\qquad
\alpha,\beta=+,-
\,.
\end{equation}
Its determinant is
\begin{equation}
\det\mathcal{G}^{\a\b}
=
\frac{4\tilde 
\mu-(1+\tilde \mu)^2}{(1-\tilde \mu)^2}
=
-1,
\end{equation}
which is compatible with the identification with a Weyl-invariant metric. Therefore, it is clear that the deformed action can be interpreted as a minimal coupling of the PCM to an auxiliary metric. This coupling to a metric is not special to the PCM, as we show in appendix~\ref{app_A0}. Notice that the second term of \eqref{Auxiliary_metric_Lag1} does not spoil this geometric interpretation. 
One might expect a density of the type
\begin{equation}
    H \sim \sqrt{-g} \,\, h(g)\,,
\end{equation}
where, in general, a two-dimensional metric will have three degrees of freedom; however, $H$ is a function of $\frak m_\pm$ only. This issue is quickly resolved as the physical fields only couple to $\frak m_\pm$ due to the scale invariance of the seed theory. Therefore, the dependence of $H$ on the third degree of freedom (the conformal factor of $g$) can be trivially integrated out. Moreover, one can identify $\frak m_\pm$ with the Beltrami differentials of the auxiliary metric, as we argue in the following subsection.
This indicates that the infinite family of deformations described by the Lagrangian \eqref{L_master} corresponds to an infinite family of metric deformations. Indeed, we are coupling the PCM to an auxiliary metric that will be fixed by the equations of motion of $\frak m_\pm$.
We can also reproduce the specific deformations that have already been studied in the literature with this approach, like $T\bar T$ \cite{Conti:2022egv}, and the mixed $T\bar T$ and root-$T\bar T$ \cite{Babaei-Aghbolagh:2024hti}. In our context, these correspond to specific choices of the ``metric self-interaction'' $H$. 
Let us now make this connection between the model \eqref{L_master} and the metric approach more precise.

\subsection{Auxiliary Beltrami metric interpretation}\label{sec_Beltrami}
We have found that the action of the PCM coupled to auxiliary fields $\frak m _\pm$ corresponds to the action of the PCM on a curved space with inverse metric density \eqref{G^{AB}}. The $\frak m_\pm$ are not simple functions of spacetime; in the auxiliary fields interpretation, we have that they are field-dependent, where the dependence is chosen by fixing the function $H(\frak m_+,\frak m_-)$ and eventually using the EOM for $\frak m_\pm$. Upon inspecting the metric density we can actually interpret these auxiliary fields as field-dependent Beltrami differentials, as we argue now.
Starting from \eqref{G^{AB}}, we have
\begin{equation}
(\mathcal{G}^{-1})_{\a\b}
=
\frac{-1}{1-\tilde \mu}
\begin{pmatrix}
2\mathfrak m_+ & 1+\tilde \mu \\
1+\tilde \mu & 2\mathfrak m_-
\end{pmatrix}.
\end{equation}
Since $(\mathcal{G}^{-1})_{\a\b}$ is Weyl-invariant, the covariant metric is
determined only up to an arbitrary local factor.
Consequently,
\begin{equation}\label{Gmet}
ds_g^2
\sim
\left(d\sigma^+ + \mathfrak m_-\,d\sigma^-\right)
\left(d\sigma^- + \mathfrak m_+\,d\sigma^+\right).
\end{equation}
The metric therefore can be written as
\begin{equation}\label{g_with_weyl}
   g_{\alpha \beta}
=
-2\Omega^2
\begin{pmatrix}
2\mathfrak m_+ & 1+\tilde \mu \\
1+\tilde \mu & 2\mathfrak m_-
\end{pmatrix},\quad \alpha,\beta=+,-
\end{equation}
where $\Omega$ is the Weyl factor.
After Euclidean continuation, defining $\tau=i\tau_E$, together with $\sigma^+=i \bar z$ and $\sigma^-=iz$, eq.~\eqref{Gmet} becomes the standard Beltrami form
\begin{equation}\label{Beltrami_metric}
ds^2
=
4\Omega^2
\left(dz+\mu_{\mathrm B}\,d\bar z\right)
\left(d\bar z+\bar\mu_{\mathrm B}\,dz\right),
\end{equation}
with the identification
\begin{equation}
\mu_{\mathrm B}
=
\mathfrak m_+ \,,
\qquad
\bar\mu_{\mathrm B}
=
\mathfrak m_-\,.
\end{equation}
This form of the metric was also used in \cite{Sakamoto:2025hwi} to define the
$T\bar T$ and root-$T\bar T$ deformations from four-dimensional Chern--Simons
theory. Our analysis shows that the discussion of \cite{Sakamoto:2025hwi}
extends to the whole family of deformations associated to an arbitrary interaction function, with the Beltrami differentials
promoted to the auxiliary fields $\mathfrak m_\pm$. 
Requiring it to be equivalent up to a diffeomorphism to the Minkowski metric in coordinates $X^\pm$,
\begin{equation}
    ds^2_g=-4 \,dX^+ d X^-\,,
\end{equation}
and reabsorbing any rescaling into $\Omega$, we find (derivatives with respect
to $\sigma^\pm$)
\begin{equation}\label{coc}
\partial_+X^+=\Omega\,,\quad \partial_-X^+=\Omega\,\mathfrak m_-\,,\qquad
\partial_-X^-=\Omega\,,\quad \partial_+X^-=\Omega\,\mathfrak m_+\,.
\end{equation}
These equations anticipate the relation with the field-dependent
coordinates of the geometric $T\bar T$ construction that we analyze in section \ref{sec_coc}. In this sense, the two charged scalar auxiliary fields can be naturally interpreted as the two metric components. The close relation between the metric interpretation we have just presented and the change of coordinates will also clarify the nature of this general family of deformations. For example, while $T\bar T$ is a change of coordinates, root-$T\bar T$ appears not to be; the appendix of \cite{Sakamoto:2025hwi} refers to it as a ``non-integrable'' change of coordinates. In our setting, these relations will allow us to identify which members of the infinite family of deformations correspond to genuine changes of coordinates. We can require that the change of coordinates is integrable by imposing $\partial_+\partial_- X^\pm= \partial_-\partial_+ X^\pm$, and the deformations satisfying this condition will produce genuine field-dependent changes of coordinates. We discuss this condition and its consequences in detail in Section~\ref{sec_coc}.

\subsection{Reduction of the massive gravity formulation}
 We start by considering the massive gravity formulation of 
\cite{Tolley:2019nmm}, extended to \cite{Babaei-Aghbolagh:2024hti}, where the matter theory is coupled to an auxiliary
zweibein $e_\mu{}^a$ and a reference zweibein $f_\mu{}^a$ through a general potential $B$ and a deformation parameter $\lambda$,
\begin{equation}\label{B_general}
S_\lambda[\phi,e,f]=S_0[\phi,e]+\frac{1}{2\lambda}\int d^2\sigma\,\det e\,B(e, f)\,.
\end{equation}
The intuition we developed in the previous section would suggest that the general potential $\det{e} \, B$ should somehow play the role of $H$.
Here $S_0[\phi,e]$ is the undeformed action for the matter fields $\phi$ coupled to the auxiliary metric
$g_{\a\b}=\eta_{ab}\,e^a_{\ \a}e^b_{\ \b}$, and the metric tensor associated to the zweibein $f_\a{}^a$ is $h_{\a\b}=\eta_{ab}\,f^a_{\ \a}f^b_{\ \b}$. The metric defined by $f$ is the physical metric of the theory, while the metric defined from $e$ is an auxiliary metric whose equations of motion enforce the deformation. Starting from this, it was shown in \cite{Babaei-Aghbolagh:2024hti} that for the $T\bar{T}$ and root-$T\bar{T}$ deformations, integrating out $e$ results in a deformed theory that lives on the curved spacetime governed by the zweibein $f$. As we are interested in relating this model to the case described by \eqref{Auxiliary_metric_Lag1}, we will focus on the case for a PCM on a flat spacetime, i.e., we fix the secondary zweibein to be flat,
\begin{equation}
f_\a{}^a=\delta_\a{}^a\,,
\end{equation}
and consider
\begin{equation}\label{B_flat}
S_\lambda[\phi,e]=S_0[\phi,e]+\frac{1}{2\lambda}\int d^2\sigma\,\,\det e\,B(e)\,.
\end{equation}
We will return to commenting on the case $f\neq\delta$ at the end of this section.

To see how this works in practice, consider the potential that reproduces the $T\bar T$ deformation \cite{Tolley:2019nmm}
\begin{equation}\label{Tolley_lag}
S_\lambda[\phi,e]
=
S_0[\phi,e]
+
\frac{1}{2\lambda}
\int d^2\sigma\,
\epsilon^{\a\b}\epsilon_{ab}
\left(e_\a{}^a-\delta_\a{}^a\right)
\left(e_\b{}^b-\delta_\b{}^b\right).
\end{equation}
Now, let us parametrize the auxiliary zweibein as follows, in order to make it compatible with the Beltrami differentials form of the metric
(as in Section~\ref{sec_Beltrami}),
\begin{equation}\label{e_param}
e_\a{}^a=\Omega
\begin{pmatrix}
1 & \mathfrak m_+\\
\mathfrak m_- & 1
\end{pmatrix},
\qquad
e^+=\Omega\left(d\sigma^++\mathfrak m_-\,d\sigma^-\right),\quad
e^-=\Omega\left(d\sigma^-+\mathfrak m_+\,d\sigma^+\right),
\end{equation}
in the $+,-$ basis. For a two-dimensional Weyl-invariant seed such as the PCM, $S_0[\Phi,e]$ depends on $\mathfrak m_\pm$ but not on the Weyl factor
$\Omega$, as the Lagrangian only couples to $\sqrt{-g} \,g^{\a \b}$. For the PCM case, we get
\begin{equation}
    S_0[j,e]=\int d^2\sigma\,\frac{
\mathfrak m_+\mathrm{tr}(j_-^2)+\mathfrak m_-\mathrm{tr}(j_+^2)-(1+\mathfrak m_-\mathfrak m_+)\mathrm{tr}(j_-j_+)
}{2
(1-\mathfrak m_-\mathfrak m_+)
}\,.
\end{equation}
On the other hand, the second term becomes,
\begin{equation}
\det(e-\delta)
=
(\Omega-1)^2-\Omega^2\tilde \mu
=
\Omega^2(1-\tilde \mu)-2\Omega+1.
\end{equation}
The Weyl factor $\Omega$ does not couple to the currents, and it appears only in the potential term $\det(e-\delta)$. Therefore, we can eliminate $\Omega$ by using its equation of motion,
\begin{equation}\label{dilaton_eom}
\Omega(1-\tilde \mu)=1,
\end{equation}
with solution
\begin{equation}
\Omega_*
=
\frac{1}{1-\tilde \mu}\,.
\end{equation}
Substituting this back into the potential term now gives
\begin{equation}
\left.
\frac{1}{\lambda}
\det(e-\delta)
\right|_{\Omega_*}
=
-\frac{\tilde \mu}{\lambda(1-\tilde\mu)}\,.
\end{equation}
Recall that $\tilde \mu=\mathfrak{m}_+\mathfrak{m}_-=\mu^2$, see eq.~\eqref{mpm to mu}, we can identify this with the interaction function of \eqref{Auxiliary_metric_Lag1} for the case of a $T\bar{T}$-deformed PCM (up to a rescaling of the parameter $\lambda$), where we have \cite{Baglioni:2025tsc}
\begin{equation}
H_{T\bar T}(\mu;\lambda)
=-
\frac{\mu^2}{2\lambda(1-\mu^2)}
\,.
\end{equation}
We can also consider the potential for a mixed $(\lambda,\gamma)$-flow, describing a mixed $T\bar T$ and $\sqrt{T\bar T}$-flow as considered in \cite{Babaei-Aghbolagh:2024hti}. For a generic zweibein $f$, the purely gravitational part of the action is built out of
\begin{equation}
\begin{split}
y_1
&=
\mathrm{tr}(e^{-1}f)
=
f_{\a}{}^a e_a{}^{\a},\quad 
y_2
=
\mathrm{tr}\left[(e^{-1}f)^2\right]
=
f_{\a}{}^ae_b{}^{\a}
f_{\b}{}^be_a{}^{\b}.
\end{split}
\end{equation}
The second term of \eqref{B_general} is then given by
\begin{equation}\label{mixed_flux}
\begin{split}
\frac{1}{2\lambda}
\int d^2\sigma\,\det e
\left(
2+y_1^2-y_2
-2y_1\cosh\frac{\gamma}{2}
+2\sqrt{2y_2-y_1^2}\,
\sinh\frac{\gamma}{2}
\right).
\end{split}
\end{equation}
Going to the flat reference frame $f=\delta $ and using the same parametrisation of $e$ as \eqref{e_param}, we find
\begin{equation}
 \det{e}\, B(e)= 2\Bigl[ \Omega^2(1-\tilde \mu)+1-2\Omega\left(\cosh{\frac\gamma2}-\sqrt{\tilde\mu}\sinh{\frac \gamma 2}\right)\Bigl]\,.
\end{equation}
Integrating out $\Omega $, we find this time
\begin{equation}
    \Omega_*=\frac{\cosh{\frac\gamma2}-\sqrt{\tilde\mu}\sinh{\frac \gamma 2}}{1-\tilde\mu}\,,
\end{equation}
which, upon substitution back into the action, gives us
\begin{equation}
    \frac{1}{2\lambda}\det e\,B(e)=-\frac{1}{2\lambda}\frac{\tilde \mu-1+\cosh{\gamma}(\tilde\mu+1)-2\sqrt{\tilde\mu}\sinh{\gamma}}{1-\tilde\mu}\,.
\end{equation}
Once more, this is in agreement with the interaction function of the mixed flow $H(\mu,\lambda,\gamma)$ that was previously found in \cite{Baglioni:2025tsc}.

Following these examples, it is clear that the conformal factor only enters the gravitational potential and is in general eliminated through its EOM. Defining the combination $\mathcal V=\det(e) B(e)$, we have
\begin{equation}\label{Omega_eom_general}
\partial_\Omega \mathcal V(\Omega,\mathfrak m_\pm)=0
\quad\Longrightarrow\quad
\Omega=\Omega_*(\mathfrak m_\pm)\,.
\end{equation}
This results in a potential term that is a function of $\mathfrak m_\pm$ only, and we obtain the action,
\begin{equation}
\begin{aligned}
\label{Auxiliary_metric_Lag}
&S_\lambda[j,e]=\int d^2\sigma\,\mathcal L_{\mathfrak m},
\\
\mathcal L_{\mathfrak m}
=\frac14\,\mathcal G^{\a\b}\,\mathrm{tr}(j_\a j_\b)
\,+&H(\mathfrak m_+,\mathfrak m_-),
\qquad
H=\frac{1}{2\lambda}\,\mathcal V\big(\Omega_*(\mathfrak m_\pm),\mathfrak m_\pm\big)\,.
\end{aligned}
\end{equation}
This is exactly the model \eqref{m_pm_metric} of the previous section. However, note that the map from $B$
to $H$ is not one-to-one, since different functions of $(\Omega,\mathfrak m_\pm)$
can give the same $H$ after $\Omega$ is integrated out.
This shows that every deformation of a conformal invariant seed of the form \eqref{B_flat} can be interpreted as an auxiliary field model of the type
\eqref{Auxiliary_metric_Lag}. In contrast, the whole family described by
\eqref{L_master} corresponds to a family of metric deformations, in which the PCM
is coupled to an auxiliary metric fixed by the equations of motion of
$\mathfrak m_\pm$.

We conclude with a comment on the more general case in which the physical two-dimensional spacetime is curved, $f\neq\delta$. It is clear from the approach of \cite{Tolley:2019nmm,Babaei-Aghbolagh:2024hti} that the reference zweibein $f$ enters the action only through its coupling to the auxiliary zweibein $e$ in the potential $B(e,f)$. 
Therefore, the curved-space case can be implemented in the action \eqref{Auxiliary_metric_Lag} by allowing $H$ to depend on the components of $f$ as well as on $\mathfrak m_\pm$.

\section{Integrability of metric deformations} \label{integrability}
In this section we focus on the Lax integrability of theories deformed via the metric approach. As argued before, these arbitrary metric couplings are closely related to the AFSM construction \cite{Ferko:2024ali}, whose integrability has been studied in detail for various seed theories. Using the PCM as an example, in section \ref{sec_1} we clarified how the so-called $\nu$-frame deformation \eqref{L_AFSM}, characterised by a set of Lie algebra valued auxiliary fields and a generic interaction function $E=E(\mathrm{tr}(v_+^2),\mathrm{tr}(v_-^2))$, can alternatively be described via the $\frak m_\pm$-frame \eqref{mpm_frame}, characterised by two scalar auxiliary fields $\mathfrak{m}_{\pm}$ and an interaction function $H=H(\mathfrak{m}_{+},\mathfrak{m}_{-})$. In section \ref{sec_3}, we further established how the latter description of the deformation can be reinterpreted as coupling the undeformed seed theory to an auxiliary metric, whose relevant degrees of freedom precisely correspond to the scalar fields $\mathfrak{m}_{\pm}$. This chain of relations implies that the integrability of metric deformed theories is simply inherited from the integrability of the auxiliary field deformations in the $\mathfrak{m}_{\pm}$-frame and $\nu$-frame for generic interaction functions $H$ and $E$, respectively.
It is interesting, however, to exploit the new auxiliary metric perspective to look at integrability from a different angle. In particular, Lax integrability requires showing that the EOM of a theory can be encoded in the flatness of a Lax connection, and one can prove that this property is preserved by any function $B(\Omega, \frak m_\pm)$, corresponding to a $H(\frak m_\pm)$, as long as the seed theory admits a Lax connection on an arbitrary curved spacetime $\Sigma$.

Suppose that the action is again given by \eqref{B_flat},
\begin{equation}\label{generic_B}
    S_\lambda[\phi]= S_0[\phi,e]+S_B[e]=S_0[\phi,e]+\frac{1}{2\lambda}\int d^2\sigma\, \det{(e)} \,B(e)\,.
\end{equation}
We follow \cite{Tolley:2019nmm,Babaei-Aghbolagh:2024hti}, but for simplicity restrict to the case where, after integrating out the auxiliary metric, the deformed physical theory is defined in flat Minkowski space. In particular, we will choose $f=\delta$. Since $e$ is an auxiliary zweibein, after imposing its equations of motion, it will depend on the physical fields $\phi$. We can write the solution for $e$ to these equations as $e^*=e^*(\phi)$, where in general the latter has dependence on the derivatives of $\phi$, which we will however keep implicit in the following. Taking the variation of the above action with respect to $\phi$ when the zweibein $e$ is on-shell, by using the functional chain rule, we can obtain
\begin{equation}\label{variation_proof}
\delta S_\lambda[\phi,e^*]
=
\int d^2x\,
\Bigg[
\frac{\delta S_0[\phi,e^*]}{\delta\phi(x)}
+
\int d^2y\,
\textcolor{black}{
\underbrace{
\textcolor{black}{
\left(
\frac{\delta S_0[\phi,e^*]}{\delta e^*(y)}
+
\frac{\delta S_B[e^*]}{\delta e^*(y)}
\right)
}
}_{\textcolor{black}{\text{vanishes for }e=e^*}}
}
\frac{\delta e^*(y)}{\delta\phi(x)}
\Bigg]
\,\delta\phi(x)\,\, .
\end{equation}
At this point, one can note that the second and third terms can be recombined into the variation of the full action with respect to the zweibein evaluated on the $e$-EOM, namely $\frac{\delta S_\lambda}{\delta e}|_{e=e^*}$, which by construction vanishes when $e$ is on-shell. Therefore, the term in the second integral vanishes when $e$ is integrated out, leaving us with
\begin{equation}\label{generic_eom}
  \delta S_\lambda[\phi, e^*]=\int d^2x\frac{\delta S_0[\phi, e^*]}{\delta \phi(x)} \delta \phi(x)\,.
\end{equation}
For the sake of clarity, we remark that while the variation appearing on the left-hand side of \eqref{generic_eom} on $S_{\lambda}$ is a total variation with respect to $\phi$, the one appearing on the right-hand side on $S_0$ is a partial variation, acting on the explicit dependence of $S_{0}$ on $\phi$ and computed with the auxiliary zweibein held fixed. Hence, the main conclusion one can draw from the above calculation is that while the on-shell zweibein $e^*$ depends on $\phi$ through its equations of motion, its implicit field dependence does not contribute to the variation of the deformed action, which is simply obtained by varying the seed action $S_0[\phi,e^*]$ with respect to $\phi$ at fixed $e^*$.
 
This implies that the EOM of the physical field $\phi$ obtained from $S_\lambda$ coincide with those obtained from an action $S_0[\phi,e^*]$, which themselves are the EOM of a seed theory on a field-dependent metric obtained by ignoring the dependence of the metric on $\phi$ when explicitly computing the variation. Therefore, as long as the seed theory admits an encoding of the EOM in the form of a flatness condition for a Lax connection on an arbitrary $\Sigma$, the EOM of $S_\lambda [\phi]$ correspond to those of the seed model in a curved geometry $e$, which is actually field-dependent, $e=e^*(\phi)$. 
The conclusion of this analysis is that one is able to define the Lax connection of metric-deformed models as the Lax connection of the respective seeds, evaluated on a curved, field-dependent spacetime $\Sigma$. Specifically, 
\begin{equation}\label{curved_Lax_general}
    \frak L^{(\lambda)}[ \phi,\eta\,;z]=\frak L^{(0)}[\phi,g\,;z]\,,
\end{equation}
where we have schematically identified $\frak L^{(\lambda)}$ as the Lax connection of a metric deformed theory. A dependence on the Minkowski metric $\eta$ has also been included, since these deformations were previously interpreted (in the sense of the auxiliary fields of Section~\ref{sec_1}) as stress-tensor deformations on flat geometries $\Sigma$. On the other hand, $\frak L^{(0)}$ in \eqref{curved_Lax_general} is the Lax of the seed theory defined on a manifold $\Sigma$ with metric $g_{\a\b}=\eta_{ab}\,{e^*}^a_{\ \a}{e^*}^b_{\ \b}$, where $e^*=e^*(\phi)$ is the solution of the $e$-EOM. 
Note that \eqref{curved_Lax_general} was a key idea in obtaining the Lax connections in \cite{Sakamoto:2025hwi} for the $T\bar T$ and root-$T\bar T$ cases. Our argument here shows how this powerful idea applies to any metric deformation. 
In the next subsection we explicitly show this identification in a couple of relevant examples, commenting on how this novel point of view on the deformation also allows us to resolve some questions raised in previous works.

\subsection{Examples of Lax connections}
Let us apply the above mechanism to some concrete examples. We argued above that the EOM of the deformed models are equivalent to the EOM of the seed models on a curved field-dependent two-dimensional metric. Here, using such a proposal, we illustrate how the recipe of \eqref{curved_Lax_general} can be used to obtain the Lax connection for a deformed theory.
The results of the previous discussion imply that the deformed Lax connections on a flat $\Sigma$ are equivalent to the undeformed connections on the curved metric \eqref{g_with_weyl}. Hence, knowing the expression for the Lax of a theory on a generic $\Sigma$ means that, using the expression of the metric in terms of the Beltrami differentials, one can write down the Lax in terms of $\frak m_\pm$ (and the physical fields $\phi)$. This expression coincides in turn with the one for the Lax of the deformed model, as long as the Beltrami differentials are promoted to auxiliary fields whose EOMs are obtained by varying \eqref{Auxiliary_metric_Lag} w.r.t. $\frak m_\pm$. This recipe provides a very general and elegant construction for the Lax connection of deformed theories. In all the cases where the $\nu$-frame construction has been applied, including those studied in appendix~\ref{app_A0}, finding the Lax required a case-by-case analysis, while this new perspective provides a model-independent geometric way of directly defining the deformed Lax connection, provided the undeformed one can be defined on an arbitrary $\Sigma$.

We will start by considering the case of the PCM, so as to make the steps more transparent, and then study the two less trivial examples of Yang-Baxter and Non-Abelian T-dual sigma models. The deformed Lax connections obtained via the coupling to \eqref{G^{AB}} will be in agreement with the ones previously obtained via coupling to the auxiliary fields in the $\nu$-frame. In particular, to compare the two expressions, we will integrate out $v_\pm$ in favor of $\mathfrak{m}_{\pm}$ from the $\nu$-frame Lax connections using \eqref{v-app}. Interestingly, studying the Yang-Baxter and Non-Abelian T-dual examples will also allow us to clarify a puzzle raised in \cite{Bielli:2026ggs} on the possibility of realising the auxiliary field deformations of such theories, which in the $\nu$-frame are characterised by what was referred to in \cite{Bielli:2026ggs} as a splitting mechanism of the second type,\footnote{This simply means that the flat and conserved current of the seed theory splits, upon deformation, into two deformed currents which are respectively flat and conserved. See section 4.3 of \cite{Bielli:2026ggs} for more on this.} via the coupling to a field-dependent metric. In the following analysis, it will be convenient to use the Hodge star definition
\begin{equation}
(\star \omega)_{\alpha}=-\sqrt{-g} \, \epsilon_{\alpha\beta}g^{\beta\gamma}\omega_{\gamma} \,\, ,
\end{equation}
which, after choosing the convention $\epsilon_{+-}=+1$, in terms of \eqref{G^{AB}} leads to
\begin{equation}\label{Hodge-mpm}
(\star \omega)_{\pm}=\pm\Biggl[\frac{1+\mathfrak{m}_{+}\mathfrak{m}_{-}}{1-\mathfrak{m}_{+}\mathfrak{m}_{-}}\omega_{\pm}-\frac{2\mathfrak{m}_{\pm}}{1-\mathfrak{m}_{+}\mathfrak{m}_{-}}\omega_{\mp}\Biggr] \,\, .
\end{equation}

\paragraph{PCM.} It is well known that the PCM admits a Lax connection on any two-dimensional curved $\Sigma$ (see \cite{Kluson:2007vw, Driezen:2021cpd} for more on this point), which takes the form
\begin{equation}
    \frak L^{(0)}=\frac{j+z\star j}{1-z^2} \,\, .
\end{equation}
In this case $j$ is the Maurer-Cartan form and one can directly exploit \eqref{Hodge-mpm} to land on
\begin{equation}\label{Lax_mpm}
    \frak L^{(0)}_\pm[j_\pm,g\,;z]=\frac{1}{1-z^2}\Bigl[j_\pm\pm z\Bigl(\frac{1+\frak m_+\frak m_-}{1-\frak m_+\frak m_-}j_\pm-\frac{2\frak m_\pm}{1-\frak m_+\frak m_-}j_\mp\Bigl)\Bigl]\,.
\end{equation}
This coincides with the Lax connection that had previously been derived from the AFSM. For example one can check it against the one presented in \cite{Baglioni:2025tsc}.\footnote{One should use \eqref{mpm to mu} to compare with the results of \cite{Baglioni:2025tsc}.}
Note that for the PCM case, the identification is almost trivial. The Lax formulation of the EOM being independent of the geometry is guaranteed because the Maurer-Cartan current is the flat as well as the conserved current here. The same is not true for other models where the Lax is built out of currents which themselves depend on the underlying geometry. Let us describe this scenario by studying the Yang-Baxter and the non-Abelian T-dual models.

\paragraph{Yang-Baxter deformations.}  We start by identifying the correct Lax for the undeformed theory on a generic $\Sigma$. Following \cite{Idiab:2024bwr}, one can write such a Lax connection as
\begin{equation}\label{Lax_YB_geom}
\frak L^{(0)}=\frac{ I + z \star  I }{1-z^2} \,\, ,
\end{equation}
with the current defined as
\begin{equation}\label{I-Yang-Baxter}
I=\frac{1}{1-\eta^2\mathcal{R}_{g}^2} \, j -\frac{\eta \mathcal{R}_{g}}{1-\eta^2\mathcal{R}_{g}^2} \, \star j \,\, .
\end{equation}
As anticipated, a key difference with the PCM is that the flat and conserved current $I$ now itself depends on the Hodge-star operator on $\Sigma$, and hence explicitly depends on its geometry deformed by \eqref{G^{AB}}. However, using the relation \eqref{Hodge-mpm} on the $\star j$ contained in $I$ as well as in $\star I$ we obtain,
\begin{equation}\label{YB_I}
\begin{aligned}
I_{\pm}=&\mathcal{O}_{+}\mathcal{O}_{-}j_{\pm}\mp M\mathcal{O}_{+}\mathcal{O}_{-}\Biggl[\frac{1+\mathfrak{m}_{+}\mathfrak{m}_{-}}{1-\mathfrak{m}_{+}\mathfrak{m}_{-}}j_{\pm}-\frac{2\mathfrak{m}_{\pm}}{1-\mathfrak{m}_{+}\mathfrak{m}_{-}}j_{\mp}\Biggr] \,\, ,
\\
(\star I)_{\pm}=& \mathcal{O}_{+}\mathcal{O}_{-}(\star j)_{\pm}-M\mathcal{O}_{+}\mathcal{O}_{-}j_{\pm} 
\\
=& \mathcal{O}_{+}\mathcal{O}_{-}\Biggl[\pm\frac{1+\mathfrak{m}_{+}\mathfrak{m}_{-}}{1-\mathfrak{m}_{+}\mathfrak{m}_{-}}j_{\pm}\mp\frac{2\mathfrak{m}_{\pm}}{1-\mathfrak{m}_{+}\mathfrak{m}_{-}}j_{\mp}\Biggr]-M\mathcal{O}_{+}\mathcal{O}_{-}j_{\pm}\,. 
\end{aligned}
\end{equation}
Now, let us compare this Lax connection with the one constructed in \cite{Bielli:2026ggs}, in the presence of auxiliary fields on a flat $\Sigma$ in lightcone coordinates, which was given by
\begin{equation}\label{Lax_yangian}
    \frak L^{(\lambda)}_{\pm}=\frac{\mathcal A_{\pm} \pm z \mathcal B_{\pm} }{1-z^2}\,.
\end{equation}
Here the respective flat and conserved currents are given by
\begin{equation}
\mathcal{A}_{\pm}=\frac{1}{1\mp\eta\mathcal{R}_{g}}\Bigl(j_{\pm}\pm \eta \mathcal{R}_{g}(2v_{\pm}) \Bigr)
\qquad \text{and} \qquad \mathcal{B}_{\pm}=-\frac{1}{1\mp\eta\mathcal{R}_{g}}\Bigl(j_{\pm}+ 2v_{\pm} \Bigr) \,\, .
\end{equation}
As suggested earlier, one can proceed by integrating out $v_{\pm}$ in favor of the scalars $\mathfrak{m}_{\pm}$ using \eqref{v-app}, with $\mathcal{K}_{\pm}=j_{\pm}$ and $\mathcal{O}_{\pm}=(1\pm M)^{-1}$ with $M=\eta \mathcal{R}_{g}$, finding that
\begin{equation}\label{AB-YB}
\begin{aligned}
\mathcal{A}_{\pm}=&\frac{1}{1-\mathfrak{m}_{+}\mathfrak{m}_{-}}\Bigl[ \mathcal{O}_{\pm}j_{\pm}-\mathfrak{m}_{+}\mathfrak{m}_{-}\mathcal{O}_{\mp}j_{\pm}\pm 2\mathfrak{m}_{\pm}\mathcal{O}_{+}M\mathcal{O}_{-}j_{\mp} \Bigr]\,\,,
\\
\mathcal{B}_{\pm}=&\frac{1}{1-\mathfrak{m}_{+}\mathfrak{m}_{-}}\Bigl[ \mathcal{O}_{\pm}j_{\pm}+\mathfrak{m}_{+}\mathfrak{m}_{-}\mathcal{O}_{\mp}j_{\pm}- 2\mathfrak{m}_{\pm}\mathcal{O}_{+}\mathcal{O}_{-}j_{\mp} \Bigr] \,\, .
\end{aligned}
\end{equation}
These currents, after a bit of rearrangement, precisely recover the expression for the current present in \eqref{Lax_YB_geom}; in particular, we observe from \eqref{YB_I} that $\mathcal{A}_\pm=I_\pm$ and $\pm\mathcal{B}_\pm=(\star I)_\pm$. This explicitly shows the matching of the YB Lax on a flat $\Sigma$ deformed via the $\mathfrak{m}_{\pm}$-frame and the Lax of the YB undeformed theory evaluated on the curved $\Sigma$ characterised by \eqref{G^{AB}}, providing a concrete non-trivial test of \eqref{curved_Lax_general}.

\paragraph{Non-Abelian T-dual.} We again follow the same
logic as for the Yang-Baxter case. We start by identifying the correct Lax for the
undeformed theory on a generic $\Sigma$. An expression for such a Lax can, for
example, be taken from equation (3.9) in \cite{Bielli:2021hzg}, where upon
redefining the spectral parameter as $\tilde{z}=\sqrt{\frac{z-1}{z+1}}$, one finds
\begin{equation}\label{T-dual-Lax-general-sigma}
\mathfrak{L}^{(0)}=\frac{\tilde{j}+z\star \tilde{j}}{1-z^2} \,\, ,
\end{equation}
with $\tilde{j}$ defined in equation (3.3) of the same work:
\begin{equation}\label{tildej-T-dual}
\tilde{j}=-\frac{1}{1-\mathrm{ad}_{X}} \mathrm{P}_{+}(\mathrm{d}X)+\frac{1}{1+\mathrm{ad}_{X}} \mathrm{P}_{-}(\mathrm{d}X)
\quad \text{with} \quad
\mathrm{P}_{\pm}:=\frac{1\pm \star}{2} \,\, .
\end{equation}
Similarly to the previous example, the current $\tilde{j}$ depends non-trivially
on the geometry of $\Sigma$, and hence explicitly on the metric \eqref{g_with_weyl}. Therefore, using again the
relation \eqref{Hodge-mpm} on the $\star\mathrm{d}X$ contained in $\tilde{j}$ as
well as in $\star\tilde{j}$, we obtain
\begin{equation}\label{T-dual_jtilde}
\begin{aligned}
\tilde{j}_{\pm}=&-\mathcal{O}_{+}M\mathcal{O}_{-}\partial_{\pm}X-\mathcal{O}_{+}\mathcal{O}_{-}(\star \mathrm{d}X)_{\pm}
\\
=& -\mathcal{O}_{+}M\mathcal{O}_{-}\partial_{\pm}X\mp\mathcal{O}_{+}\mathcal{O}_{-}\Biggl[\frac{1+\mathfrak{m}_{+}\mathfrak{m}_{-}}{1-\mathfrak{m}_{+}\mathfrak{m}_{-}}\partial_{\pm}X-\frac{2\mathfrak{m}_{\pm}}{1-\mathfrak{m}_{+}\mathfrak{m}_{-}}\partial_{\mp}X\Biggr] 
\\
(\star \tilde{j})_{\pm}=&-\mathcal{O}_{+}\mathcal{O}_{-}\partial_{\pm}X-\mathcal{O}_{+}M\mathcal{O}_{-}(\star \mathrm{d}X)_{\pm} \,\, ,\\
=& -\mathcal{O}_{+}\mathcal{O}_{-}\partial_{\pm}X\mp\mathcal{O}_{+}M\mathcal{O}_{-}\Biggl[\frac{1+\mathfrak{m}_{+}\mathfrak{m}_{-}}{1-\mathfrak{m}_{+}\mathfrak{m}_{-}}\partial_{\pm}X-\frac{2\mathfrak{m}_{\pm}}{1-\mathfrak{m}_{+}\mathfrak{m}_{-}}\partial_{\mp}X\Biggr] \,\, ,
\end{aligned}
\end{equation}
where $\mathcal{O}_{\pm}=(1\pm M)^{-1}$ and $M=\mathrm{ad}_{X}$. Now, let us compare this Lax connection with the one constructed in
\cite{Bielli:2026ggs} in the context of auxiliary fields. The Lax connection for the $\nu$-frame deformed model on  flat
$\Sigma$ in lightcone coordinates reads
\begin{equation}\label{Lax-T-dual-def}
\mathfrak{L}^{(\lambda)}_{\pm}=\frac{\mathcal{A}_{\pm}\pm z\mathcal{B}_{\pm}}{1-z^2}
\,\,,
\end{equation}
with the flat and conserved currents now given by
\begin{equation}
\mathcal{A}_{\pm}=\mp\frac{1}{1\mp\mathrm{ad}_{X}}\Bigl(\partial_{\pm}X\pm 2v_{\pm} \Bigr)
\qquad \text{and} \qquad
\mathcal{B}_{\pm}=\mp\frac{1}{1\mp \mathrm{ad}_{X}}\Bigl(\partial_{\pm}X +\mathrm{ad}_{X}(2v_{\pm}) \Bigr) \,\, .
\end{equation}
Once again, using \eqref{v-app} to trade
$v_{\pm}$ for $\mathfrak{m}_{\pm}$, leads to\footnote{More precisely, the expressions \eqref{AB-T-dual} are obtained starting from $\mathcal{A}_{\pm}$ and $\mathcal{B}_{\pm}$ in \cite{Bielli:2026ggs} by first redefining $v_{\pm}\rightarrow\pm v_{\pm}$, substituting then $v_{\pm}$ using \eqref{v-app} with $\mathcal{K}_{\pm}=\partial_{\pm}X$ and finally replacing $X\rightarrow-X$. This chain of substitutions ensures consistency between appendix \ref{app_A0} and the T-dual model originally constructed in \cite{Bielli:2024fnp}, making the conventions more suitable for this work, as compared to the ones previously chosen in \cite{Bielli:2025uiv}.}
\begin{equation}\label{AB-T-dual}
\begin{aligned}
\mathcal{A}_{\pm}=&\frac{1}{1-\mathfrak{m}_{+}\mathfrak{m}_{-}}\Bigl[\mp\mathcal{O}_\mp\partial_{\pm}X \mp \mathfrak{m}_{+}\mathfrak{m}_{-}\mathcal{O}_{\pm}\partial_{\pm}X\pm 2\mathfrak{m}_{\pm}\mathcal{O}_{+}\mathcal{O}_{-}\partial_{\mp}X \Bigr]\,\,,
\\
\mathcal{B}_{\pm}=&\frac{1}{1-\mathfrak{m}_{+}\mathfrak{m}_{-}}\Bigl[\mp\mathcal{O}_\mp\partial_{\pm}X \pm \mathfrak{m}_{+}\mathfrak{m}_{-}\mathcal{O}_{\pm}\partial_{\pm}X+ 2\mathfrak{m}_{\pm}M\mathcal{O}_{+}\mathcal{O}_{-}\partial_{\mp}X \Bigr] \,\, .
\end{aligned}
\end{equation}
These currents, after a bit of rearrangement, precisely recover the expressions
obtained from the undeformed Lax, and we observe from \eqref{T-dual_jtilde} that
$\mathcal{A}_{\pm}=\tilde{j}_{\pm}$ and
$\pm\mathcal{B}_{\pm}=(\star\tilde{j})_{\pm}$. Similarly, this explicitly shows the matching between the T-dual
Lax on a flat $\Sigma$ deformed via the $\mathfrak{m}_{\pm}$-frame and the
undeformed Lax of the same theory evaluated on a $\Sigma$ with the structure
\eqref{G^{AB}}, providing yet another non-trivial test of 
\eqref{curved_Lax_general}.

To close this example-based subsection, we would finally like to stress how the above results not only provide an explicit test of the relation \eqref{curved_Lax_general}, but also lead to a resolution of a puzzling feature observed in the last paragraph of section 4.3 in \cite{Bielli:2026ggs}. There, it was observed that the existence of two different splitting mechanisms, induced by the $\nu$-frame deformation, seemed to prevent the possibility of a field-dependent metric interpretation for theories characterised by the second type of mechanism, where the flat and conserved current of the seed theory is non-trivially deformed into two separate currents. From that point of view, the field-dependent metric interpretation seemed to be prevented in such theories due to the fact that, in the expression for the Lax, only one of the two deformed currents is acted upon by a Hodge star operator, in which the deformation could be reabsorbed. The above analysis shows however that this apparent obstruction was a mere artifact of the flat lightcone coordinates, which effectively hid a more complicated Hodge dependence appearing throughout the Lax and the currents themselves, as seen explicitly via the examples \eqref{I-Yang-Baxter} and \eqref{tildej-T-dual}.

The examples presented here may also be viewed within the broader formalism of the $\mathcal{E}$-model, which we plan to address in upcoming work
\cite{WIP1}. This will also address the recently developed $\lambda$-model in
the $\nu$-frame \cite{Ferko:2026Lambda} within the framework established here, and provide a
universal approach to integrability. In particular, since the $\mathcal{E}$-model is formulated at the level of the
Poisson structure \cite{Klimcik:1995dy,Klimcik:1996nq,Lacroix:2020flf,Severa:2017kcs}, it should provide a natural setting in which to investigate
strong integrability for all of these models together, going beyond the
existence of a flat Lax connection established here.

\subsection{Comments on some previous field-dependent metric results}\label{sec_comments}

Despite its simplicity, equation \eqref{generic_eom} offers a useful perspective on several results from the literature. Specifically, although the connection between the AFSM and the metric approach was somewhat expected, this new point of view allows to explain various previously obtained results. One example, as already mentioned, is the interpretation of certain deformed Lax connections in \cite{Bielli:2026ggs} as couplings to a non-trivial $2d$ metric. In hindsight, it is now obvious how, in that case, the missing ingredient leading to an interpretation of the AFSM \cite{Ferko:2024ali} as a metric deformation was the lack of a manifest coupling to the $2d$ metric on $\Sigma$. This is indeed obscured by the $\nu$-frame, described in terms of Lie-algebra-valued auxiliary fields $v_\pm$, while it is made much more clear and direct when using the $\frak m_\pm$-frame, characterised by two simpler scalar auxiliary fields. This is also connected to the observation that the family of deformations studied in this work is equivalent, at the level of EOM, to coupling the physical fields to a {field-dependent metric}. This fact was observed in \cite{Ferko:2025bhv}, where the AFSM on the $\mathbb{CP}^{N-1}$ target space was studied, and in \cite{Ferko:2024yhc}, with a focus on theories of electrodynamics in $d=4$. The formulation presented here helps understanding the relation between these interpretations. In particular, following \cite{Ferko:2025bhv}, the EOM of the AFSM \eqref{L_AFSM} on $\mathbb{CP}^{N-1}$ can be written as
\begin{equation}\label{field_dep_eoms}
\mathcal D_\mu\Bigl((h^{-1})^{\mu \nu}j^{(2)}_\nu\Bigl)=0\,,
\end{equation}
where $h$ is a field-dependent metric that can be expressed in terms of $v_\pm$ and $E$. The $\frak m_\pm$ scalar frame introduced in section \ref{sec_1} makes it clear that this field-dependent metric $h$ corresponds to $g_{\alpha \beta}=\eta_{a b}\,e^a_{\,\alpha}e^b_{\,\beta}$ used in this work, after integrating out $e$ (namely $\frak m_\pm$) from \eqref{Auxiliary_metric_Lag}. The EOM \eqref{field_dep_eoms} are precisely of the form observed in \eqref{generic_eom}. They are the equations of motion on a curved spacetime $\Sigma$ where $h$ is promoted to be field-dependent. It was also observed in \cite{Ferko:2025bhv}, as well as in \cite{Ferko:2024yhc} in the context of four-dimensional non-linear electrodynamics, that such a field-dependent metric cannot simply be coupled to the model at the level of the action
\begin{equation}
    S_h=\int d^2x \,\frac12 (h^{-1})^{\mu \nu}\mathrm{tr}(j^{(2)}_\nu j^{(2)}_\mu)\,,
\end{equation}
since varying with respect to $j^{(2)}_\mu$ would not simply lead to the EOM \eqref{field_dep_eoms}, as an additional term would be generated due to the dependence of $(h^{-1})^{\mu \nu}$ on $\phi$. This extra term, which appears by naively coupling the metric at the level of the action, precisely corresponds to the contribution $ \frac{\delta S_0[\phi, e^*]}{\delta e^*(y)} \frac{\delta e^*(y)}{\delta \phi(x)}$ in \eqref{variation_proof}. On the other hand, starting from an action of the form \eqref{generic_B}, the $S_B$ term leads to a cancellation of the unwanted contribution in the EOM of $j$ due to the EOM of $e$. This makes it clear that all of these field-dependent metrics originate from a coupling to an auxiliary metric like \eqref{generic_B}. In this sense, the coupling to the auxiliary metric at the level of the action can be regarded as the fundamental reason why these field-dependent metrics appear.

\section{Auxiliary metrics vs changes of coordinates} \label{sec_coc}
We now discuss the question of whether these auxiliary deformations are (field-dependent) changes of coordinates. The auxiliary field approach allows us to impose the condition that the deformation corresponds to a change of coordinates, leading to constraints on $H$ that distinguish deformations admitting such an interpretation from those that do not. In terms of the Beltrami variables and the Weyl factor, the condition for which a frame is a change of coordinates translates to \eqref{coc} satisfying symmetry under the interchange of the second derivatives,
\begin{equation}
    \frac{\partial^2 X^{\pm}}{\partial\sigma^- \partial \sigma^{+}}=\frac{\partial ^2X^{\pm}}{\partial\sigma^+ \partial \sigma^{-}}\,.
\end{equation}
In terms of the Beltrami coordinates, this can be written as,
\begin{equation}\label{integrability_coc}
    (\partial_--\frak m_-\partial_+)\Omega=\Omega \,\partial_+\frak m_-\,,\qquad(\partial_+-\frak m_+\partial_-)\Omega=\Omega \,\partial_-\frak m_+\,.
\end{equation}
 These equations contain the conformal factor $\Omega$; however, we are interested in theories where this factor can be integrated out trivially, and it does not couple to the physical fields. As mentioned in the previous section, upon integrating out $\Omega$ from the action \eqref{Tolley_lag}, we obtain $\Omega=\frac1{1-\frak m_+\frak m_-}$. This result was observed for the specific case of the $T\bar T$ coupling. However, it
extends to a broader class of potentials. For a potential of
the form $\mathcal V(\Omega,\frak m_\pm)=\Omega^2(1-\tilde\mu)\,B(\Omega,\tilde\mu)$
and for the zweibein solution $ \Omega_*=\frac{1}{1-\tilde\mu}\,$, the problem
reduces to finding the interaction function such that $H(\tilde\mu)=\mathcal V(\Omega_*,\frak m_+,\frak m_-)$. Conversely, for any given $H(\tilde\mu)$ one can always choose a $B$ such that
$\Omega_*$ solves the equation of motion for $\Omega$ and reproduces the interaction function H. The same holds for the mixed flow described by \eqref{mixed_flux}, where starting from
a different function $B(e)$, we still obtain the same reduced potential
$H(\tilde\mu)$ upon integrating out $\Omega$. In the rest of this section we therefore take the zweibein solution as $\Omega=\frac{1}{1-\frak m_+\frak m_-}$ and find the interaction function $H(\tilde\mu)$ that corresponds to a genuine field-dependent change-of-coordinates type of deformation. 

Under the assumptions described above, equation \eqref{integrability_coc} then becomes
\begin{equation}
(\partial_--\frak m_-\partial_+)\frac1{1-\frak m_+\frak m_-}= \frac{\partial_+\frak m_-}{1-\frak m_+\frak m_-}\,,\qquad
    (\partial_+-\frak m_+\partial_-)\frac1{1-\frak m_+\frak m_-}=\frac{\partial_-\frak m_+}{1-\frak m_+\frak m_-}\,.
\end{equation} 
Now substituting $\partial_-\frak m_+$ from the second equation into the first equation and $\partial_+\frak m_-$ from the first into the second, we obtain
\begin{equation}\label{integrability_cond}
    \partial_+\frak m_-=\frak m_+\partial_-\frak m_-\,,\quad \text{and} \quad \partial_-\frak m_+=\frak m_-\partial_+\frak m_+\,.
\end{equation}
For a usual change of coordinates, $\frak m_\pm$ are functions of $(\sigma^+,\sigma^-)$. Here, $\frak m_\pm$ are auxiliary fields, and we require \eqref{integrability_cond} to be satisfied when both $\frak m_\pm$ and the physical fields are taken on-shell.
 As explained in \cite{Conti:2019dxg}, the symmetry of the second derivatives is a consequence of the conservation of the stress-energy tensor (see appendix \ref{app_A} for the derivation). Now, we use the stress-energy tensor \eqref{stress_energy_mupm} and the EOM to check whether \eqref{integrability_cond} is satisfied. We present the main steps of the calculation here, leaving the details to Appendix \ref{app_2}.
 
The first step is to show that the $\frak m_\pm$ EOM \eqref{mupm_eom} can be written as
\begin{equation}\label{mupm_eom_stress_tens}
  H_{\frak m_\pm}+ \frac{\mathrm{tr}(j_\mp^2)(1+\frak m_+\frak m_-)-2\frak m_\mp\mathrm{tr}(j_+j_-)}{2(1-\frak m_+\frak m_-)^2}=0\,,
\end{equation}
where $H_{\mathfrak {m_\pm}}=
\partial_{\mathfrak {m_\pm}}H$. Now we see that the second term is simply the component of the stress-energy tensor in \eqref{stress_energy_mupm}:
\begin{equation}\label{H_stress_energy}
  H_{\frak m_+}(1-\frak m_+\frak m_-)-T_{--}=0\,\qquad H_{\frak m_-}(1-\frak m_+\frak m_-)-T_{++}=0\,.
\end{equation}
This identification is true when the auxiliary fields are on-shell. The EOM of the matter fields imply the conservation of the stress-energy tensor. Therefore, taking a derivative of \eqref{H_stress_energy} and using the conservation equation $\partial^\mu T_{\mu \nu}=0$, we obtain
\begin{equation}
    \partial_\pm(H_{\frak m_\pm}(1-\frak m_+\frak m_-))=-\partial_\mp T_{+-}.
\end{equation}
Furthermore, using $T_{+-}=-H$, we obtain
\begin{equation}\label{H_ode_rel}
    \partial_\pm(H_{\frak m_\pm}(1-\frak m_+\frak m_-))=\partial_\mp H\,.
\end{equation}
Until now we have used the equations of motion of both the auxiliary and the matter fields to impose two PDEs \eqref{H_ode_rel} in terms of two derivatives and two variables: $\partial_\pm \frak m_\pm$. For any choice of $H$ (as long as it depends on the product $\tilde \mu=\frak m_+\frak m_-$) we can, in principle, solve these PDEs to find, say, $\partial_+\frak m_+$ and $\partial_-\frak m_-$, in terms of the other two derivatives.

Alternatively, we can also use \eqref{H_ode_rel} differently, by requiring that the
change of coordinates condition \eqref{integrability_cond} be satisfied, which fixes $\partial_+\frak m_-$
and $\partial_-\frak m_+$ in terms of $\partial_-\frak m_-$ and
$\partial_+\frak m_+$. Therefore, we see that this requirement imposes an ODE on $H(\tilde{\mu})$ through \eqref{H_ode_rel}, which amounts to
\begin{equation}
2 H_{\tilde \mu}-(1-\tilde \mu)H_{{\tilde \mu}{\tilde \mu}}=0,
 \end{equation}
 where $H_{\tilde \mu}=\frac{d H}{d\tilde{\mu}}$ and $H_{\tilde \mu\tilde \mu}=\frac{d^2 H}{d\tilde{\mu}^2}$. The above equation is a second-order differential equation that can be reduced to first order since $H$ only appears through $H_{\tilde \mu}$. The general solution is 
 \begin{equation}\label{ODE_sol}
     H(\tilde \mu)=\frac{k_1}{1-\tilde \mu}+k_2\,,
 \end{equation}
 for some constants $k_1$ and $k_2$.
 The solution \eqref{ODE_sol} can be easily seen to correspond to the $T\bar T$ interaction function
 \begin{equation}
     H_{T\bar T}=-\frac{\tilde \mu}{2\lambda(1-\tilde \mu)},
 \end{equation} 
 up to a constant. We therefore conclude that, within the Lorentz-invariant family $H=H(\tilde\mu)$ with the prescription $\Omega=(1-\tilde\mu)^{-1}$, $T\bar T$ is the unique deformation for which the chosen coframe is locally a change of coordinates on-shell.

\subsection{Comparison with the field-dependent change of coordinates}

It is instructive to compare the present construction with the field-dependent change of coordinates of \cite{Conti:2018tca}. A central observation of that work is that the classical equations of motion of a $T\bar T$-deformed theory can be related to those of the undeformed theory by a field-dependent change of coordinates. The $T\bar T$ deformation is interpreted geometrically as a dynamical redefinition of the worldsheet coordinates. For example, a free boson obeys $\partial_w\partial_{\bar w}\phi=0$ in the $\mathbf w$ coordinates, while in the $\mathbf z$ coordinates it obeys the $T\bar T$-deformed equations of motion. The relation between the two coordinate systems is encoded in the Jacobian of the transformation, whose entries are determined by components of the stress-energy tensor of the undeformed theory,
\begin{equation}\label{jacobian_tateo}
\mathcal{J}
=
\begin{pmatrix}
\partial_z w & \partial_z \bar{w} \\
\partial_{\bar z} w & \partial_{\bar z} \bar{w}
\end{pmatrix}
=
\frac{1}{\Delta(\mathbf{w})}
\begin{pmatrix}
1+2\tau\,\Theta_{0}(\mathbf{w})
&
-2\tau\,{\rm T}_{2}(\mathbf{w})
\\[2mm]
-2\tau\,\bar{\rm T}_{2}(\mathbf{w})
&
1+2\tau\,\bar{\Theta}_{0}(\mathbf{w})
\end{pmatrix}.
\end{equation}
The inverse Jacobian gives the derivatives of the deformed coordinates with respect to the undeformed coordinates,
\begin{equation}
\mathcal{J}^{-1}
=
\begin{pmatrix}
\partial_{w}z & \partial_{w}\bar{z} \\
\partial_{\bar{w}}z & \partial_{\bar{w}}\bar{z}
\end{pmatrix}
=
\begin{pmatrix}
1+2\tau\,\bar{\Theta}_{0}(\mathbf{w})
&
2\tau\,{\rm T}_{2}(\mathbf{w})
\\[2mm]
2\tau\,\bar{\rm T}_{2}(\mathbf{w})
&
1+2\tau\,\Theta_{0}(\mathbf{w})
\end{pmatrix},
\end{equation}
where $\Delta(\mathbf{w})$ is
\begin{equation}
\Delta(\mathbf{w})
=
\bigl(1+2\tau\,\Theta_{0}(\mathbf{w})\bigr)
\bigl(1+2\tau\,\bar{\Theta}_{0}(\mathbf{w})\bigr)
-4\tau^{2}\,{\rm T}_{2}(\mathbf{w})\,
\bar{\rm T}_{2}(\mathbf{w}).
\end{equation}
Here, $\tau$ is the deformation parameter, and ${\rm T}_2(\mathbf w)=T_{ww}$ is the holomorphic component of the stress-energy tensor, $\Theta_0(\mathbf w)$ is its trace component, and $\bar{\rm T}_2$, $\bar\Theta_0$ are the corresponding antiholomorphic components, and all the components of the stress-energy tensor are evaluated in the undeformed coordinates $\mathbf w$.

 Translating this now into our conventions, we work with the coordinates $\sigma^\pm$, which correspond to the deformed coordinates, and the undeformed coordinates obey \eqref{coc}. Going to Minkowski signature, we can identify $X^\pm$ with $\mathbf w$ and $\sigma^\pm$ with $\mathbf z$. We can use the same notation as in \cite{Conti:2019dxg} and write
\begin{equation}\label{jacobian_mupm}
\mathcal{J}
=
\begin{pmatrix}
\partial_{z^+}w^+ & \partial_{z^+} w^-\\
\partial_{z^-} w^+ & \partial_{z^-} w^-
\end{pmatrix}
=
\frac{1}{1-\frak m_+\frak m_-}
\begin{pmatrix}
1
&
\frak m_+
\\[2mm]
\frak m_-
&
1
\end{pmatrix}.
\end{equation} 
However, note that $\frak m_\pm$ in this matrix are functions of the deformed coordinates $\mathbf z$, not $\mathbf w$. Comparing with \eqref{jacobian_tateo}, and using the fact that the seed theory is conformal and hence the undeformed theory will not have a $\Theta(\mathbf{w})$ contribution, we can read off the identification,
\begin{equation}
    \frak m_\pm=-2\tau T_{\pm\pm}(\mathbf{w})=\tau \,\mathrm{tr}(j_\pm j_\pm)(\mathbf{w})\,,
\end{equation}
where we have used that the undeformed $T_{\pm\pm}$ corresponds to $T^{(PCM)}_{\pm\pm}$. The right-hand side depends on the undeformed coordinates, and we can use the inverse Jacobian to express it in the deformed ones. For instance,
\begin{equation}
\begin{split}
    \frak m_+=&\,\tau \Bigl(\frac{\partial z^+}{\partial w^+}\frac{\partial z^+}{\partial w^+} \mathrm{tr}(j_+ j_+)+2\frac{\partial z^+}{\partial w^+}\frac{\partial z^-}{\partial w^+} \mathrm{tr}(j_+ j_-)+\frac{\partial z^-}{\partial w^+}\frac{\partial z^-}{\partial w^+} \mathrm{tr}(j_- j_-)\Bigl)\\
    =&\,\tau\Bigl(\mathrm{tr}(j_+ j_+)-2 \frak m_+ \mathrm{tr}(j_+ j_-)+ \frak m_+^2 \mathrm{tr}(j_- j_-)\Bigl)\,,
\end{split}
\end{equation}
where now the right-hand side is expressed in terms of $\mathbf{z}$. To avoid cumbersome notation, from here on, unless specified otherwise, all quantities are understood to be functions of the deformed coordinates. Now using the on-shell conditions \eqref{relativistic_relation_app} we find
\begin{equation}
    \frak m_+=\tau \Bigl( \mathrm{tr}(j_+j_+)(1+\frak m_+\frak m_-)-2\frak m_+ \mathrm{tr}(j_+j_-)\Bigl)\,.
\end{equation}
Furthermore, using the definition of the stress-energy tensor \eqref{stress_energy_mupm} we get
\begin{equation}
    \frak m_+= -2\tau\,(1-\frak m_+\frak m_-)\, T_{++}\,,
\end{equation}
while $\frak m_-$ obeys analogous relations. For $H^{(T \bar T)}=-\frac{\frak m_+\frak m_-}{2\lambda(1-\frak m_+\frak m_-)}$, we then obtain
\begin{equation}
-    \frac{\frak m_+}{2\tau(1-\frak m_+\frak m_-)}=H^{(T\bar T)}_{\frak m_-}(1-\frak m_+\frak m_-)=T_{++}\,.
\end{equation}
Up to identifying $\lambda=\tau$, this is exactly what we had obtained from the auxiliary equations of motion \eqref{H_stress_energy}. Hence, this shows that the Jacobian \eqref{jacobian_mupm} is exactly analogous to what was found in \cite{Conti:2018tca} for the dynamical changes of coordinates.

 \section{Conclusions and outlooks}

In this work we have developed a charged scalar auxiliary field formulation of a broad class of two-dimensional stress-tensor, $\TTbar$-like deformations and related it to auxiliary metric constructions. The starting point is a master action containing both the Lie algebra-valued, $\nu$-frame, AFSM auxiliaries $v_\pm$ and two Lorentz-charged scalar auxiliary fields $\mathfrak m_\pm$. Integrating out either set of the auxiliary fields provides a direct relation between the usual AFSM description \cite{Ferko:2024ali} and the scalar $\mathfrak m_\pm$-frame, extending the one scalar $\mu$-frame developed in~\cite{Baglioni:2025tsc}. These relations between the various auxiliary descriptions, including the CH approach, are summarized in Table~\ref{tab:intro-frames}. In particular, the two charged scalars naturally describe deformations that need not preserve Lorentz invariance, while the Lorentz invariant sector is recovered when the interaction depends only on $\tilde\mu=\mathfrak m_+\mathfrak m_-$.

The central geometric result is that the auxiliary fields $\mathfrak m_\pm$ may be identified with the Beltrami differentials of an auxiliary worldsheet metric. For classically Weyl-invariant seed theories, the conformal factor decouples from the matter sector and can be eliminated algebraically from the gravitational potential, leaving precisely the scalar interaction $H(\mathfrak m_+,\mathfrak m_-)$. In this way, the scalar auxiliary field description provides a bridge between AFSMs and the auxiliary-metric formulations of $T\bar T$-like deformations established in~\cite{Tolley:2019nmm} and subsequently generalised in~\cite{Tsolakidis:2024wut,Babaei-Aghbolagh:2024hti}. A consequence of this correspondence is that for this subclass of seed theories any metric self-interaction preserves the integrability of the deformed theory.

\begin{table}[!t]
\centering
\begingroup
\renewcommand\baselinestretch{1.05}\small
\renewcommand{\arraystretch}{1.12}
\setlength{\tabcolsep}{4pt}
\begin{tabular}{@{}>{\raggedright\arraybackslash}p{0.23\textwidth}>{\raggedright\arraybackslash}p{0.28\textwidth}>{\raggedright\arraybackslash}p{0.45\textwidth}@{}}
\toprule
Description & Auxiliary data & Relation between formulations \\
\midrule
Master Lagrangian & $v_\pm$, $\mu_{2\pm}$, interaction $H$ & Describes both frames. \\[0.3cm]

Vector AFSM ($\nu$-frame) & $v_\pm$, interaction $E$ & Eliminate $\mu_{2\pm}$ from the master action. \\
Charged scalar frame & $\mathfrak m_\pm=2\mu_{2\pm}$, interaction $H$ & Eliminate $v_\pm$; identify $\mathfrak m_\pm$ with Beltrami data. \\
Auxiliary geometry & $e_\alpha{}^a$, reference $f_\alpha{}^a$ & For Weyl-invariant matter, eliminating the conformal factor gives the scalar auxiliary potential $H$. \\
Neutral scalar / CH & $\mu^2=\tilde{\mu}=\mathfrak m_+\mathfrak m_-$ & In the Lorentz-invariant sector, eliminate the charged ratio; $y=(1+\mu)/(1-\mu)$ connects to Russo--Townsend and Courant--Hilbert variables. \\
\bottomrule
\end{tabular}
\caption{Auxiliary descriptions of the stress-tensor sector. Equivalences are understood locally on regular algebraic branches; the neutral scalar and Courant--Hilbert (CH) descriptions concern the Lorentz invariant subfamily.}
\label{tab:intro-frames}
\endgroup
\end{table}

As an extension of the analysis of \cite{Sakamoto:2025hwi}, our formulation also gives a simple geometric criterion for classical integrability. Whenever the seed theory admits a Lax representation of its equations of motion on arbitrary curved two-dimensional spacetimes, coupling it to an algebraic auxiliary metric with a matter-independent potential preserves the corresponding Lax connection. The deformed equations may therefore be viewed as the seed equations evaluated on a field-dependent solution for the auxiliary geometry.

Our results also clarify the relation between auxiliary metrics and field-dependent coordinate transformations. These notions are not equivalent in general. Within the Lorentz invariant family $H=H(\tilde\mu)$ and the zweibein prescription $\Omega=(1-\tilde\mu)^{-1}$ considered in this work, imposing the local
integrability conditions for the coframe yields
\[
H(\tilde\mu)=\frac{k_1}{1-\tilde\mu}+k_2
\]
as the only solution. Up to normalisation and an additive constant, this is precisely the $T\bar T$ interaction. Thus, the larger family of integrability-preserving auxiliary metric deformations does not in general admit a field-dependent coordinate interpretation.\footnote{Other irrelevant deformations involving conserved currents, such as $J\bar T$~\cite{Guica:2017lia}, may also admit field-dependent coordinate interpretations, but they lie outside the purely stress-tensor family studied here.}

Another natural question is whether the auxiliary field formulation can make manifest some of the symmetry-preserving properties that are already known for the $T\bar T$ deformation. For conformal seed theories, the classical $T\bar T$ deformation retains an infinite-dimensional, field-dependent realisation of conformal symmetry, with the corresponding charges forming two commuting copies of the Witt algebra~\cite{Guica:2020uhm}. Supersymmetry likewise was also shown to be preserved by $\TTbar$ and $J\bar{T}/T\bar{J}$ deformations, see~\cite{Baggio:2018rpv,Chang:2018dge,Jiang:2019hux,Chang:2019kiu,Jiang:2019trm}. It would be interesting to understand how these symmetry properties emerge directly in the various auxiliary field descriptions, including the auxiliary metric formulation, and to determine to what extent they survive for general choices of the auxiliary potential $H$. Such an analysis could provide additional criteria, beyond Lax integrability, for identifying distinguished deformations within the larger family studied here.

Several other directions follow naturally from these results. First, the Beltrami and auxiliary metric interpretation suggests extending the four-dimensional Chern--Simons and $\mathcal E$-model \cite{Klimcik:1995dy,Klimcik:1996nq,Severa:2017kcs} constructions of integrable field theories~\cite{Lacroix:2020flf,Sakamoto:2025hwi} to the larger class of auxiliary potentials considered here. This direction is currently being explored in ongoing work~\cite{WIP1}, where we aim to develop a master action-like formulation for the $\mathcal E$-model framework and investigate its integrability properties, along with its relation to the Beltrami description with a view towards connecting with the four-dimensional Chern--Simons construction. This may also help clarify the scope and interpretation of the construction beyond the class of integrable conformal seed theories studied in this work. Second, the master action suggests higher-spin analogues in which the quadratic chiral invariants are supplemented by higher-degree ones and coupled to additional charged scalar auxiliaries. These theories are not expected, in general, to admit an ordinary metric interpretation and should instead involve a richer auxiliary, higher-spin geometry.

It would also be interesting to explore which aspects of the construction persist for non-conformal or massive integrable seed theories.
In that case, the Weyl factor no longer decouples from the matter sector, so the two-scalar $\mathfrak m_\pm$ description is no longer sufficient, and the full auxiliary geometry must be retained. It is still not clear if seed theories with potentials (e.g., sinh-Gordon) admit a non-trivial family of stress-tensor deformations. It could be that the integrability of this family of metric deformations is in general broken for any $B(\Omega,\frak m_\pm)$ except for $B_{T\bar T}$. $T\bar T$ could be the exception due to being a change of coordinates. We could also look at this from a different angle. The existence of these deformations is related to the integrability of the seed theory on a generic spacetime. Therefore, the curved-spacetime Lax criterion developed here provides a natural starting point for investigating whether analogous integrability-preserving deformations exist in this more general setting. Finally, the appearance of field-dependent metrics in four-dimensional duality invariant nonlinear electrodynamics \cite{Ferko:2024yhc} and previous works on the connection with massive gravity \cite{Floss:2023nod} suggest that an analogous auxiliary gravity interpretation may exist there as well. We leave these questions for future work.

\medskip
\section*{Acknowledgements}

We are grateful to Mattia Cesàro, Lewis Cole, Christian Ferko, Michele Galli, Ben Hoare, Jue Hou, Zejun Huang, Sergei Kuzenko, Tommaso Morone, Liam Smith, Dmitri Sorokin, and Roberto Tateo for discussions and collaboration on topics related to this project.
We acknowledge the language models ChatGPT-5.6 and ChatGPT-6 used during this research.
The authors take full responsibility for the preparation of this manuscript and for the accuracy of this work, having reviewed, verified, and approved all AI-generated content.
N.\,B. is supported by a postgraduate scholarship at the University of Queensland. 
D.\,B. has been supported by a Young Scientist Training (YST) Fellowship from Asia Pacific Center for Theoretical Physics (APCTP).
D.\,B. and P.\,S. have been supported by a UKRI Future Leaders Fellowship Renewal (grant number UKRI2067). 
G.\,T.-M. and N.\,B. have been supported by the Australian Research Council (ARC) Discovery Project DP240101409, and a faculty start-up funding of UQ’s School of Mathematics and Physics.

\appendix
\section{Extension of metric coupling beyond the PCM}\label{app_A0}
In this appendix we describe how the common structure \eqref{mpm_frame_general} of the $\mathfrak{m}_{\pm}$-frame Lagrangian emerges for various classes of auxiliary field sigma models sharing the $\nu$-frame description of \cite{Bielli:2025uiv}\footnote{This common picture includes the PCM, T-dual, SSSM and (bi)-YB deformations, even though in this manuscript we will not discuss the bi-YB case.}. This can again be interpreted as a coupling to an auxiliary metric. The common point of view on such models was also exploited in \cite{Baglioni:2025tsc} to construct the respective $\mu$-frame Lagrangians, which can in turn also be recovered from the $\mathfrak{m}_{\pm}$ picture. We start by considering the following generalisation of the PCM master Lagrangian \eqref{L_master}:
\begin{equation}\label{L_master_app}
\begin{aligned}
\!\mathcal{L}_{\mathrm M}
\!\!=\!\!
\tfrac{1}{2}  \mathrm{tr}[(\mathcal{K}_{-}\!+\!2v_{-})\mathcal{O}_{-}(\mathcal{K}_{+}\!+\!2v_{+})]\!-\! \mathrm{tr}[v_{+}v_{-}]\!+\!\mu_{2-}\operatorname{tr}[v_+^2]
\!+\!\mu_{2+}\operatorname{tr}[v_-^2]
\!+\!H(\mu_{2+},\mu_{2-}),
\end{aligned}
\end{equation}
where $\mathcal{O}_{\pm}:\mathfrak{g}\rightarrow \mathfrak{g}$ are defined via $M:\mathfrak{g}\rightarrow \mathfrak{g}$ as 
\begin{equation}\label{Opm}
\mathcal{O}_{\pm}:=\frac{1}{1\pm M}
\qquad \text{with} \qquad M^{T}=-M
\qquad \text{such that} \qquad
\mathcal{O}_{\pm}^{T}=\mathcal{O}_{\mp} \,\, .
\end{equation}
As a preliminary step, we highlight that \eqref{L_master_app} describes the coupling of the physical fields, encoded in $\mathcal{K}_{\pm}$, to the $\nu$-frame auxiliary fields $v_{\pm}$ and its structure is dictated by the requirement that integrating out $v_{\pm}$ when $\mu_{2\pm}=0$ should reproduce the undeformed $2d$ seed theory
\begin{equation}
\mathcal{L}^{(0)}:=-\tfrac{1}{2}\, \mathrm{tr}[\mathcal{K}_{+}\mathcal{O}_{-}\mathcal{K}_{-}] \,\, .
\end{equation}
Equivalently, as already noted for the PCM, the structure of the first line in \eqref{L_master_app} is fixed by demanding that integrating out $\mu_{2\pm}$ recovers the $\nu$-frame Lagrangian. Independently of this perspective, one can write \eqref{L_master_app} as
\begin{equation}\label{L_M_intermediate_app}
\mathcal{L}_{\text{M}}=\tilde{\mathcal{L}}^{(0)}+\mathcal{L}^{(1)}+\mu_{2-}\operatorname{tr}[v_+^2]
+\mu_{2+}\operatorname{tr}[v_-^2]
+H(\mu_{2+},\mu_{2-}) \,\, ,
\end{equation}
with 
\begin{equation}
\begin{aligned}
\tilde{\mathcal{L}}^{(0)}=&\tfrac{1}{2} \, \mathrm{tr}[\mathcal{K}_{+}\mathcal{O}_{+}\mathcal{K}_{-}]\,\, ,
\\
\mathcal{L}^{(1)}=&\tfrac{1}{2} \, \mathrm{tr}[2v_{-}\mathcal{O}_{-}\mathcal{K}_{+}+2v_{+}\mathcal{O}_{+}\mathcal{K}_{-}+4v_{-}\mathcal{O}_{-}v_{+}-2v_{+}v_{-}] \,\, ,
\end{aligned}
\end{equation}
noting that $\tilde{\mathcal{L}}^{(0)}$ simply differs from $\mathcal{L}^{(0)}$ by a minus sign and transposition of the inserted operator. In particular, one can easily recognise that such transposition implies
\begin{equation}\label{L0-tildeL0-decomposition}
\begin{aligned}
\mathcal{L}^{(0)}\!=\!&-\tfrac{1}{2}\mathrm{tr}[\mathcal{K}_{+}\mathcal{O}_{-}\mathcal{K}_{-}]\!=\!-\tfrac{1}{2}\mathrm{tr}[\mathcal{K}_{+}\mathcal{O}_{+}\mathcal{O}_{-}\mathcal{K}_{-}]-\tfrac{1}{2}\mathrm{tr}[\mathcal{K}_{+}\mathcal{O}_{+}M\mathcal{O}_{-}\mathcal{K}_{-}]\!=\!\mathcal{L}_{G}^{(0)}+\mathcal{L}_{B}^{(0)}\,\, ,
\\
\tilde{\mathcal{L}}^{(0)}\!=\!&\tfrac{1}{2}\mathrm{tr}[\mathcal{K}_{+}\mathcal{O}_{+}\mathcal{K}_{-}]\!=\!\tfrac{1}{2}\mathrm{tr}[\mathcal{K}_{+}\mathcal{O}_{+}\mathcal{O}_{-}\mathcal{K}_{-}]-\tfrac{1}{2}\mathrm{tr}[\mathcal{K}_{+}\mathcal{O}_{+}M\mathcal{O}_{-}\mathcal{K}_{-}] \!=\!-\mathcal{L}_{G}^{(0)}+\mathcal{L}_{B}^{(0)}\,\, ,
\end{aligned}
\end{equation}
namely that $\mathcal{L}^{(0)}$ and $\tilde{\mathcal{L}}^{(0)}$ simply differ by the sign of their symmetric part, associated to the metric of the sigma model. This feature will play a role below, when reinterpreting the result of integrating out $v_{\pm}$ from \eqref{L_M_intermediate_app} in terms of a coupling of the undeformed theory to a metric parametrised by $\mu_{2\pm}$. From \eqref{L_M_intermediate_app}, the EOM for $v_{\pm}$ when $\mu_{2\pm}=0$ read
\begin{equation}
v_{\pm}\deq(1-2\mathcal{O}_{\mp})^{-1} \mathcal{O}_{\mp}\mathcal{K}_{\pm}\,=-\mathcal O_\pm \mathcal K_\pm,
\end{equation}
and the requirement
\begin{equation}
\tilde{\mathcal{L}}^{(0)}+\mathcal{L}^{(1)}\deq \mathcal{L}^{(0)}
\qquad \text{translates into} \qquad
\mathcal{O}_{+}+\mathcal{O}_{-}=2\mathcal{O}_{+}\mathcal{O}_{-} \,\, ,
\end{equation}
which is indeed respected by \eqref{Opm}. At this point we proceed in deriving the structure \eqref{mpm_frame_general} by varying $v_{\pm}$ in \eqref{L_master_app} and obtaining, after some manipulations, that
\begin{equation}
\Bigl[(1-2\mathcal{O}_{+})(1-2\mathcal{O}_{-})-4\mu_{2+}\mu_{2-} \Bigr]v_{\pm}=(1-2\mathcal{O}_{\pm})\mathcal{O}_{\mp}\mathcal{K}_{\pm}+2\mu_{\pm 2}\mathcal{O}_{\pm}\mathcal{K}_{\mp} \,\, .
\end{equation}
While one could in principle already substitute back into the master Lagrangian \eqref{L_master_app}, it is convenient to first notice that for the operators in \eqref{Opm} one has
\begin{equation}
(1-2\mathcal{O}_{+})(1-2\mathcal{O}_{-})=1 \,\, ,
\end{equation}
which considerably simplifies the rearrangements, eventually leading to 
\begin{equation}\label{v-app}
    v_\pm=-\frac{\mathcal O_\pm(\mathcal K_\pm-2\mu_{2\pm} \mathcal K_\mp)}{1-4\mu_{2+}\mu_{2-}}\,.
\end{equation}
In particular, substituting \eqref{v-app} into \eqref{L_master_app} one finds three independent structures
\begin{equation}
\mathrm{tr}[\mathcal{K}_{-}\mathcal{O}_{+}\mathcal{O}_{-}\mathcal{K}_{-}] \quad , \quad 
\mathrm{tr}[\mathcal{K}_{+}\mathcal{O}_{+}\mathcal{O}_{-}\mathcal{K}_{+}]
\quad , \quad
\mathrm{tr}[\mathcal{K}_{-}\mathcal{O}_{+}\mathcal{O}_{-}\mathcal{K}_{+}] \,\, ,
\end{equation}
and after using the decomposition \eqref{L0-tildeL0-decomposition} for $\tilde{\mathcal{L}}^{(0)}$, one arrives at 
\begin{equation}\label{L_M_final_appendix}
\begin{aligned}
\mathcal{L}_{\mathfrak{m}}=\frac{1}{2(1-4\mu_{+2}\mu_{-2})}\Biggl( &
-(1+4\mu_{+2}\mu_{-2})\mathrm{tr}[\mathcal{K}_{-}\mathcal{O}_{+}\mathcal{O}_{-}\mathcal{K}_{+}]
\\
&+2\mu_{+2}\mathrm{tr}[\mathcal{O}_{-}\mathcal{K}_{-}\mathcal{O}_{-}\mathcal{K}_{-}]+2\mu_{-2}\mathrm{tr}[\mathcal{O}_{+}\mathcal{K}_{+}\mathcal{O}_{+}\mathcal{K}_{+}]\Biggr) 
\\
&- \tfrac{1}{2} \, \mathrm{tr}[\mathcal{K}_{+}\mathcal{O}_{+}M\mathcal{O}_{-}\mathcal{K}_{-}]+H(\mu_{+2},\mu_{-2}) \,\, .
\end{aligned}
\end{equation}
At this point one can proceed, in analogy with the PCM, by rewriting 
\begin{align}
\mathcal{L}^{(0)}(\eta)+H=&-\tfrac{1}{2}\mathrm{tr}[\mathcal{K}_{+}\mathcal{O}_{-}\mathcal{K}_{-}]+H(\mu_{+2},\mu_{-2})
\\
=&\tfrac{1}{4}\sqrt{-\det{(\eta_{\gamma\delta})}}\,\eta^{\alpha\beta} \mathrm{tr}[\mathcal{K}_{\alpha}\mathcal{O}_{+}\mathcal{O}_{-}\mathcal{K}_{\beta}]-\tfrac{1}{2}\mathrm{tr}[\mathcal{K}_{+}\mathcal{O}_{+}M\mathcal{O}_{-}\mathcal{K}_{-}] +H(\mu_{+2},\mu_{-2})\,\, ,
\notag
\end{align}
where we used again $\eta^{+-}=\eta^{-+}=-\tfrac{1}{2}$ and $\eta_{+-}=\eta_{-+}=-2$. Using the standard definition of the Hilbert stress-energy tensor on the seed Lagrangian $\mathcal{L}^{(0)}$, one then finds
\begin{equation}\label{T_ab}
T_{\alpha\beta}^{(0)}=-\frac{2}{\sqrt{-\text{det}(\eta_{\gamma\delta})}}\frac{\delta \mathcal{L}^{(0)}}{\delta \eta^{\alpha\beta}} \,\, =-\tfrac{1}{2}\mathrm{tr}[\mathcal{K}_{\alpha}\mathcal{O}_{+}\mathcal{O}_{-}\mathcal{K}_{\beta}]+\tfrac{1}{4}\eta_{\alpha\beta}\eta^{\gamma\delta}\mathrm{tr}[\mathcal{K}_{\gamma}\mathcal{O}_{+}\mathcal{O}_{-}\mathcal{K}_{\delta}] \,\, ,
\end{equation}
which allows to realise that, in terms of $\frak m_\pm=2\mu_{2\pm}$, \eqref{L_M_final_appendix} can be rearranged as
\begin{equation}\label{Lm-rearranged-app}
\begin{aligned}
\mathcal{L}_{\mathfrak{m}}=\frac{1}{(1-\mathfrak{m}_{+}\mathfrak{m}_{-})} \Bigl[ (1+\mathfrak{m}_{+}\mathfrak{m}_{-})\mathcal{L}_{G}^{(0)}-\mathfrak{m}_{+}T_{--}^{(0)}-\mathfrak{m}_{-}T_{++}^{(0)} \Bigr]+ \mathcal{L}_{B}^{(0)}\!+\!H(\frak m_+,\frak m_-)  \, ,
\end{aligned}
\end{equation}
recovering the expression \eqref{mpm_frame_general}.
Finally, replacing $\sqrt{-\det{(\eta_{\gamma\delta})}}\,\eta^{\alpha\beta}$ with the metric data in \eqref{G^{AB}} precisely allows to obtain that
\begin{equation}
\!\!\mathcal{L}^{(0)}\!(g)\!+\!H\!=\!\tfrac{1}{4}\!\sqrt{\!-\!\det{\!(g_{\rho\sigma}\!)}}\,g^{\mu\nu} \mathrm{tr}[\mathcal{K}_{\mu}\mathcal{O}_{+}\mathcal{O}_{-}\mathcal{K}_{\nu}]\!-\!\tfrac{1}{2}\mathrm{tr}[\mathcal{K}_{+}\mathcal{O}_{+}M\mathcal{O}_{-}\mathcal{K}_{-}]\!+\!H(\mu_{+2},\mu_{-2}) \!=\!\mathcal{L}_{\mathfrak{m}} ,
\end{equation}
with $\mathcal{L}_{\mathfrak{m}}$ explicitly given in \eqref{L_M_final_appendix}, or alternatively \eqref{Lm-rearranged-app}. Hence, this shows that the metric-deformation interpretation extends to the class of models described by \eqref{L_master_app}.

\section{Stress-energy tensor derivation}\label{app_A} 
In this appendix we present the calculation of the stress-energy tensor for the family of deformations of the PCM.
The Lagrangian we start with is:
\begin{equation}\label{PCM_mupm}
\mathcal{L}
=
\frac{
 \frak m_+\mathrm{tr}(j_-^2)+ \frak m_-\mathrm{tr}(j_+^2)-(1+ \frak m_-\frak m_+)\mathrm{tr}(j_-j_+)
}{
2(1-\frak m_-\frak m_+)
}
+
H(\frak m_+,\frak m_-).
\end{equation}
We use Noether's procedure since the model is in general not relativistic and it is not immediately obvious how to couple it to a metric.
It is immediate to see that under global spacetime translations $\delta \sigma^\a=a^\a$ the Maurer-Cartan currents change like
\begin{equation}
    \delta j_\a=a^{\b}\partial_\b j_\a\,,
\end{equation}
meaning that the variation of the Lagrangian under said translations will be
\begin{equation}
    \delta \mathcal L'= \frac{\partial \mathcal L'}{\partial j_\a}\delta j_\a=a^\b \frac{\partial \mathcal L'}{\partial j_\a} \partial_\b j_\a \,.
\end{equation}
Here we are working inside the trace, meaning that $\mathcal L $ is meant to be $\mathcal L=\mathrm{tr}(\mathcal L')$. We have ignored the auxiliary fields as their contribution vanishes automatically on-shell.
At the same time, since the action is also a scalar under translations, it will vary by $\delta \mathcal L'=a^{\b}\partial_\b \mathcal L'$. Now, using the Maurer-Cartan equation, we find:
\begin{equation}
    a^{\b}\partial_\b \mathcal L'= a^\b \frac{\partial \mathcal L'}{\partial j_\a} \partial_\b j_\a = a^\b \frac{\partial \mathcal L'}{\partial j_\a} \partial_\a j_\b + a^\b \frac{\partial \mathcal L'}{\partial j_\a} [j_\a,j_\b]\,,
\end{equation}
and the last term will vanish because of the properties of the trace. Therefore, we are left with the usual 
\begin{equation}
    a^{\b}\partial_\a\left(\delta^\a_\b \mathcal L'-\frac{\partial \mathcal L'}{\partial j_\a}  j_\b\right)=-a^\b \partial_\a\left(\frac{\partial \mathcal L'}{\partial j_\a} \right)j_\b\,=\,0\,,
\end{equation}
where the right-hand side vanishes since it corresponds to the Euler-Lagrange equations. This means that $T^{\a}_{\,\,\,\b}=\mathrm{tr}(\frac{\partial \mathcal L'}{\partial j_\a} j_\b)-\delta^\a_\b \mathcal L$.
Hence, the components of the energy-momentum tensor will be, using a normalisation compatible with \eqref{T_ab}:
\begin{equation}\label{stress_energy_mupm}
\begin{split}
&T^+_{\,\,\,-}=-\frac{\frak m_-\mathrm{tr}(j_+j_-)}{2(1-\frak m_-\frak m_+)}+\frac{(1+ \frak m_+\frak m_-)\mathrm{tr}(j_-^2)}{4(1-\frak m_-\frak m_+)}\;, \,\,\, T^+_{\,\,\,+}=-\frac14\frac{\frak m_- \mathrm{tr}(j_+^2)-\frak m_+\mathrm{tr}(j_-^2)}{1-\frak m_-\frak m_+}+\frac12H\;,\\
 &T^-_{\,\,\,+}=-\frac{\frak m_+\mathrm{tr}(j_+j_-)}{2(1-\frak m_-\frak m_+)}+\frac{(1+ \frak m_+\frak m_-)\mathrm{tr}(j_+^2)}{4(1-\frak m_-\frak m_+)}\;, \,\,\, T^-_{\,\,\,-}=-\frac14\frac{\frak m_+ \mathrm{tr}(j_-^2)-\frak m_-\mathrm{tr}(j_+^2)}{1-\frak m_-\frak m_+}+\frac12H\;.\\
\end{split}
\end{equation}
As expected, we are able to write the stress-energy tensor components only in terms of auxiliary objects which will implicitly depend on the physical fields through their equations of motion. Indeed, if we use the EOM of $\frak m_\pm$ we get:
\begin{align}\label{stress_energy_mupm_onshell}
    &T^+_{\quad-}=-\frac12\left(H_{\frak m_+}-\frak m_-^2 H_{\frak m_-}\right)\;, \quad T^+_{\quad+}=-\frac12\left(H_{\frak m_+}\frak m_+-H_{\frak m_-}\frak m_-\right)+\frac12H\;,\\
 &T^-_{\quad+}=-\frac12\left(H_{\frak m_-}-\frak m_+^2 H_{\frak m_+}\right)\;, \quad T^-_{\quad-}=-\frac12\left(H_{\frak m_-}\frak m_-+H_{\frak m_+}\frak m_+\right)+\frac12H\;.
\end{align}
Notice that Lorentz-invariant deformations are those where 
\begin{equation}\label{rel_H}
    H_{\frak m_-}\frak m_--H_{\frak m_+}\frak m_+=0,
\end{equation}
or equivalently $H=H(\frak m_+ \frak m_-)$.

\section{Conditions for $H$ to be a change of coordinates}\label{app_2}

In this section we explain in detail the calculation presented in section \ref{sec_coc}.
For a general deformation, we can inspect the equations of motion for the auxiliaries and find\footnote{Notice that we have a singularity at $\frak m_+\frak m_-=1$. This corresponds to what was already noticed in \cite{Ferko:2025bhv} and one could see it as the point corresponding to the $\gamma\to \infty$ limit of $\sqrt{\TTbar}$.}
\begin{equation}\label{mupm_eom}
    \frac{\partial \mathcal L}{\partial \frak m_\pm}=0=\frac{2H_{\frak m_\pm}(1-\frak m_+\frak m_-)^2+\mathrm{tr}(j_\mp^2)+\mathrm{tr}(j_\pm^2)\frak m_\mp^2-2\frak m_\mp\mathrm{tr}(j_+j_-)}{2(1-\frak m_+\frak m_-)^2}\,.
\end{equation}
Notice that for general relativistic interaction functions $H$ we have from \eqref{rel_H}, $H_{\frak m_+}\frak m_+=H_{\frak m_-}\frak m_-$ meaning that we can combine the two equations of \eqref{mupm_eom} to find
\begin{equation}\label{relativistic_relation_app}
    \mathrm{tr}(j_+^2)\frak m_-=\mathrm{tr}(j_-^2)\frak m_+\,.
\end{equation}
Thus \eqref{mupm_eom} becomes
\begin{equation}\label{mupm_eom_stress_tens_app}
  H_{\frak m_\pm}+ \frac{\mathrm{tr}(j_\mp^2)(1+\frak m_+\frak m_-)-2\frak m_\mp\mathrm{tr}(j_+j_-)}{2(1-\frak m_+\frak m_-)^2}=0\,.
\end{equation}
Now we see that the second term is just the component of the stress-energy tensor in \eqref{stress_energy_mupm}
\begin{equation}\label{H_stress_energy_app}
  H_{\frak m_+}(1-\frak m_+\frak m_-)+2T^+_{\quad-}=0\,\qquad H_{\frak m_-}(1-\frak m_+\frak m_-)+2T^-_{\quad+}=0\,.
\end{equation}
To reiterate, this identification works when the auxiliary fields are on-shell. Then, taking the matter fields also on-shell implies the conservation of the energy-momentum tensor. We can take a derivative of \eqref{H_stress_energy} and use the conservation equation $\partial^\mu T_{\mu \nu}=0$ to obtain
\begin{equation}
    \partial_\pm(H_{\frak m_\pm}(1-\frak m_+\frak m_-))=2\partial_\mp T^\mp_{\quad\mp}
\end{equation}
and then use the results of \eqref{stress_energy_mupm_onshell} to obtain
\begin{equation}\label{H_ode_rel_app}
    \partial_\pm(H_{\frak m_\pm}(1-\frak m_+\frak m_-))=\partial_\mp H\,.
\end{equation}
We can rewrite these differential equations as equations for $\partial_\pm\frak m_\mp$ as
 \begin{equation}\label{eq_abc_app}
 \begin{split}
     A\,\partial_+\frak m_-&= B \,\frak m_+\partial_-\frak m_-+ C\, \frak m_-^2 \partial_+\frak m_+\,,\\
     A\,\partial_-\frak m_+&= B \,\frak m_-\partial_+\frak m_++ C\, \frak m_+^2 \partial_-\frak m_-\,,
\end{split}
 \end{equation}
 where we have 
\begin{equation}
 \begin{split}
     A&= (1-4{\tilde \mu}) H_{\tilde \mu}^2 +2{\tilde \mu}(1-2{\tilde \mu}) H_{\tilde \mu} H_{{\tilde \mu}{\tilde \mu}}+{\tilde \mu}^2(1-{\tilde \mu})H_{{\tilde \mu}{\tilde \mu}}^2\,,\\
     B&= H_{\tilde \mu}^2\,,\\
     C&= (H_{\tilde \mu}+{\tilde \mu} H_{{\tilde \mu}{\tilde \mu}})(2 H_{\tilde \mu} -(1-{\tilde \mu}) H_{{\tilde \mu}{\tilde \mu}})\,,\\
\end{split}
 \end{equation}
with the subscripts indicating derivatives, $H_{\tilde \mu}=\frac{dH}{d{\tilde \mu}}$.
 Now we can impose that the deformation is a change of coordinates by imposing \eqref{integrability_cond}. If we do so, we find equations
 \begin{equation}\label{ABC}
 \begin{split}
     (A-B)\,\partial_+\frak m_-&= C\, \frak m_- \partial_-\frak m_+\,,\\
     (A-B)\,\partial_-\frak m_+&= C\, \frak m_+ \partial_+\frak m_-\,.
\end{split}
 \end{equation}
 Notice already that if $A-B$ vanishes but $C$ does not (or vice versa) both $\partial_\pm\mathfrak m_\mp$ are forced to be zero.
 We can then solve for both derivatives, finding:
 \begin{equation}
     (\partial_-\frak m_+)\frac{(A-B)^2-\tilde \mu C^2}{A-B}=0\,, \quad \text{and} \quad (\partial_+\frak m_-)\frac{(A-B)^2-\tilde \mu C^2}{A-B}=0\,.
 \end{equation}
 Imposing that the numerator $(A-B)^2-\tilde{\mu} C^2$ vanishes, gives us the following differential equation
 \begin{equation}\label{de for H}
        \tilde \mu\left(2 H_{\tilde \mu}-(1-\tilde \mu)H_{{\tilde \mu}{\tilde \mu}}\right)^2\Bigl((1-4{\tilde \mu})H_{\tilde \mu}^2+2{\tilde \mu}(1-2{\tilde \mu})H_{\tilde \mu} H_{{\tilde \mu}{\tilde \mu}}+(1-{\tilde \mu}){\tilde \mu}^2 H_{{\tilde \mu}{\tilde \mu}}^2\Bigl)=0.
 \end{equation}
 These are second-order differential equations that can be just reduced to first order since $H$ only appears derived. The first factor of \eqref{de for H} has a general solution,
 \begin{equation}\label{PDE_sol}
 H(\tilde \mu)=\frac{k_1}{1-\tilde \mu}+k_2\,,
 \end{equation}
for some integration constants $k_1$ and $k_2$. Notice that this is also the only solution for which both $A-B=0$ and $C=0$, which is compatible with what we had before. This solution is precisely the $T\bar T$ interaction function up to a freedom of adding an immaterial constant to the action.
The second factor of \eqref{de for H} admits two additional branches,
\begin{equation}
     H(\tilde \mu)=\Bigl(\frac{2}{1\pm\sqrt{\tilde \mu}}-4 \tanh^{-1}{(1\pm2\sqrt{\tilde \mu})}\Bigl)c_1+c_2\,,
 \end{equation}
 for some integration constants $c_1$ and $c_2$.
Notice that the limit $\tilde \mu\to 0$ diverges due to the arctanh. Hence this solution does not admit a limit to the undeformed theory. This means that \eqref{PDE_sol} is the only allowed solution. Notice that the mixed $(\lambda,\gamma)$ deformation corresponding to both $T\bar T$ and root-$T\bar T$ also does not satisfy the differential equation for $\gamma\neq0$.

\bibliographystyle{utphys}
\bibliography{master}
\end{document}